\documentclass[aps,pra,reprint,floatfix,nofootinbib,superscriptaddress,longbibliography]{revtex4-1}

\usepackage{amsmath,amsthm,amsfonts,amssymb,bm}
\usepackage{graphicx,epstopdf,nomencl,upgreek,url}
\usepackage{times}
\usepackage[colorlinks={true},pdftex]{hyperref}  
\usepackage{subfigure}
\usepackage{color}
\hypersetup{
  colorlinks,
  citecolor=blue,
  linkcolor=blue,
  urlcolor=blue}

\newcommand{\beq}{\begin{equation}}
\newcommand{\eeq}{\end{equation}}
\newcommand{\bqa}{\begin{eqnarray}}
\newcommand{\eqa}{\end{eqnarray}}
\newcommand{\bqan}{\begin{eqnarray*}}
\newcommand{\eqan}{\end{eqnarray*}}
\newcommand{\nn}{\nonumber}

\newcommand{\ket}[1]{\ensuremath{\left\vert{#1}\right\rangle}}

\newcommand{\expect}[1]{\ensuremath{\left\langle{#1}\right\rangle}}

\newcommand{\erf}[1]{Eq.~(\ref{#1})}

\newcommand{\etal}{\emph{et al.}}

\newcommand{\dg}{^{\dagger}}

\newcommand{\SNLCA}{Sandia National Laboratories, Livermore, CA 94550, USA}
\newcommand{\SNLNM}{Sandia National Laboratories, Albuquerque, NM 87123, USA}
\newcommand{\Stanford}{Edward L. Ginzton Laboratory, Stanford University, Stanford, CA 94305, USA}
\newcommand{\Waterloo}{Institute of Quantum Computing, University of Waterloo, N2L 3G1 Waterloo, Canada}

\begin{document}

\title{Self-referenced continuous-variable quantum key distribution protocol}
\author{Daniel B. S. Soh} 
\affiliation{\SNLCA}
\affiliation{\Stanford}
\author{Constantin Brif}
\affiliation{\SNLCA}
\author{Patrick J. Coles}
\affiliation{\Waterloo}
\author{Norbert\nobreak~L\"{u}tkenhaus}
\affiliation{\Waterloo}
\author{Ryan M. Camacho}
\affiliation{\SNLNM}
\author{Junji Urayama}
\affiliation{\SNLNM}
\author{Mohan Sarovar}
\email[Corresponding authors:~]{dbsoh@sandia.gov}
\email{mnsarov@sandia.gov}
\affiliation{\SNLCA}

\begin{abstract}
We introduce a new continuous-variable quantum key distribution (CV-QKD) protocol, \emph{self-referenced} CV-QKD, that eliminates the need for transmission of a high-power local oscillator between the communicating parties. In this protocol, each signal pulse is accompanied by a reference pulse (or a pair of twin reference pulses), used to align Alice's and Bob's measurement bases. The method of phase estimation and compensation based on the reference pulse measurement can be viewed as a quantum analog of intradyne detection used in classical coherent communication, which extracts the phase information from the modulated signal. We present a proof-of-principle, fiber-based experimental demonstration of the protocol and quantify the expected secret key rates by expressing them in terms of experimental parameters. Our analysis of the secret key rate fully takes into account the inherent uncertainty associated with the quantum nature of the reference pulse(s) and quantifies the limit at which the theoretical key rate approaches that of the respective conventional protocol that requires local oscillator transmission. The self-referenced protocol greatly simplifies the hardware required for CV-QKD, especially for potential integrated photonics implementations of transmitters and receivers, with minimum sacrifice of performance. As such, it provides a pathway towards scalable integrated CV-QKD transceivers, a vital step towards large-scale QKD networks.
\end{abstract}


\maketitle

\section{Introduction}
\label{sec:intro}

Quantum key distribution (QKD), which enables the generation of secure shared randomness between two distant parties (Alice and Bob)~\cite{bennett1984quantum}, is the most advanced quantum technology to date~\cite{Scarani:2009wq, Fehr.FoundPhys.40.494.2010, Lo:2014ex}. Discrete-variable QKD (DV-QKD) is the term for well-established protocols that involve generation and detection of extremely weak pulses of light (ideally, single photons). Unfortunately, significant technological challenges still remain in generation and detection of single photons, although important advances have been made over past three decades~\cite{Alleaume:2014fk, Lo:2014ex}. Protocols for an alternative approach, continuous-variable QKD (CV-QKD), were developed more recently~\cite{Scarani:2009wq, Weedbrook:2012tz}. CV-QKD utilizes conjugate continuous degrees of freedom (field quadratures) of a light pulse prepared in a Gaussian (coherent or squeezed) state to transmit the signals that constitute the shared randomness. At the receiver, the quadratures are measured using shot-noise limited balanced homodyne or heterodyne detectors, which have the advantage of not requiring single photon detection and operating at extremely high detection rates (on the order of~GHz). In particular, the coherent-state CV-QKD protocol has received much attention because of its promise of achieving information-theoretically secure key distribution with modest technological resources~\cite{Ralph:1999wo, Grosshans:520452, Grosshans2003}. The technical ease of CV-QKD is balanced by more complex and less efficient post-processing schemes for distilling a shared secret key from the imperfect shared randomness established during the quantum signal exchange portion of the protocol. However, with the recent development of higher efficiency error correction codes~\cite{jouguet2013experimental, jouguet2014high, Jouguet:2014ff} and more comprehensive security proofs~\cite{Furrer:2012dz, Furrer:2014dz, Leverrier:2015tr} for CV-QKD, it is becoming an attractive alternative to DV-QKD. A particular reason for the appeal is the expectation that the integrated photonics implementation of CV-QKD will be easier than that of DV-QKD, and such implementations are critical for the next phase of QKD development that is focused on practicality and wide-spread utilization.

A major obstacle to the implementation of CV-QKD, especially in integrated photonics, is the requirement for transmission of a local oscillator (LO) between Alice and Bob. Current fiber-based implementations co-transmit the LO with the signal states using techniques that involve combinations of time-division multiplexing (TDM), wavelength-division multiplexing (WDM), and polarization encoding~\cite{qi2007experimental, jouguet2013experimental}. Free-space implementations of CV-QKD also multiplex using the polarization degree of freedom \cite{Heim:2014fi}. Since the LO intensity dictates the quality of quadrature measurement at Bob's receiver, it is desirable to transmit a high-power LO that is many orders of magnitude more intense than the signal pulse. Due to this power disparity, multiplexing has to significantly separate the two components in order to minimize the contamination of the signal states by photons scattered from the LO (for example, this is the reason for combining polarization encoding with TDM, as in Ref.~\cite{jouguet2013experimental}). This degree of separation in multiplexing (and associated demultiplexing at Bob's receiver) greatly complicates the hardware required for CV-QKD, and is even a roadblock for integrated photonics implementations of CV-QKD since TDM and polarization manipulation and maintenance are more difficult on-chip \cite{Okamoto.LPR.2011, Bogaerts.OE.2007, Sacher.OE.2014}. Another complication associated with the requirement of LO transmission is that a relative phase shift arises between the signal and LO due to the path separation during demultiplexing at the receiver \cite{qi2007experimental, lodewyck2007quantum}. This shift can be compensated by precise calibration of the separated paths, which is however not a robust solution, or by dynamic phase estimation at the receiver, in which case the speed at which this estimation can be done becomes a practical limitation on the rate of key generation.

\begin{figure}[t]
	\centering
	\includegraphics[width=1.0\linewidth]{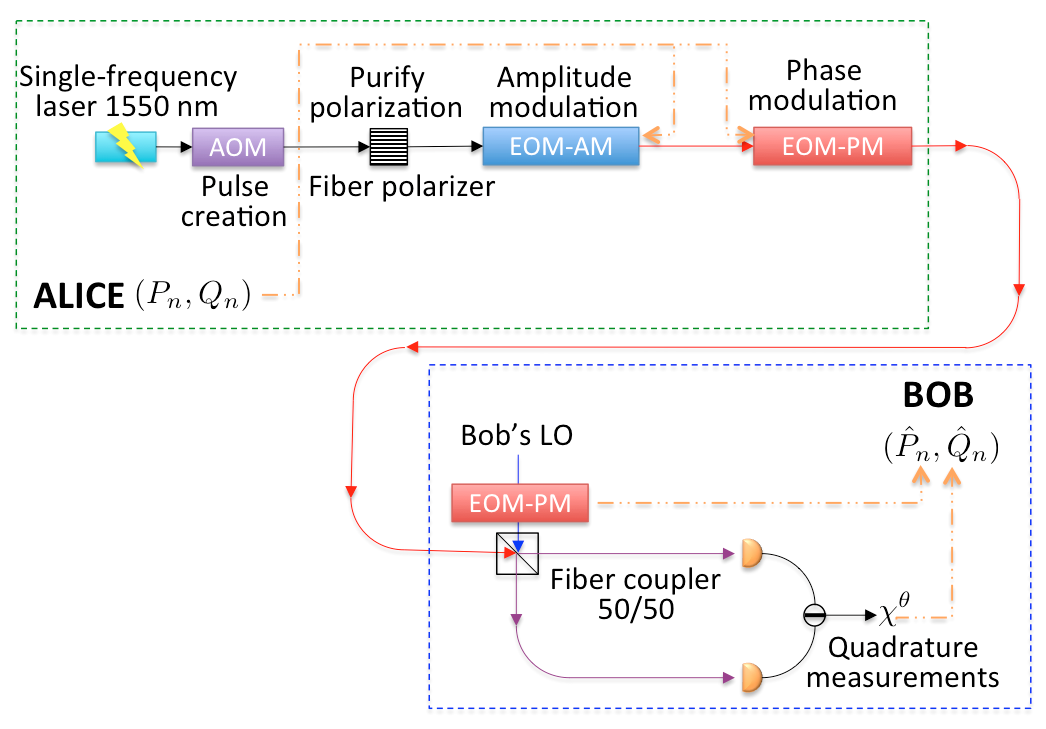}
	\caption{Hardware schematic for the SR-CV-QKD protocol. In contrast to conventional CV-QKD implementations (e.g., Ref.~\cite{jouguet2013experimental}), the hardware requirements are dramatically simplified due to elimination of LO transmission.}
	\label{fig:SR-CV-QKD}
\end{figure}

In this work, we eliminate all of the issues outlined above by developing a coherent-state CV-QKD protocol that eliminates the transmission of an LO between Alice and Bob. We achieve this by noticing that a common reference frame between Alice and Bob can be established by a method that, instead of transmitting the LO, uses regularly spaced \emph{reference pulses} whose quadratures are measured by Bob to estimate Alice's phase reference. This new protocol, which we call \emph{self-referenced} CV-QKD (SR-CV-QKD), greatly simplifies the hardware requirements at Alice's and Bob's stations since it enables them both to employ independent (truly local) LOs. In addition, SR-CV-QKD obviates a key assumption of most CV-QKD security proofs \cite{Furrer:2014dz} --- namely that the LO is trusted --- and thus provides a more secure implementation of CV-QKD. We demonstrate the key elements of SR-CV-QKD using a fiber-based setup utilizing fiber-pigtailed bulk-optics components. However, we stress that this protocol is manifestly compatible with chip-scale implementation since it only requires (low-loss and low-noise) classical optical communication components, as outlined in Fig.~\ref{fig:SR-CV-QKD}.

The remainder of the paper is organized as follows. The SR-CV-QKD protocol is described in Sec.~\ref{sec:SR-CV-QKD}. In Sec.~\ref{sec:theory} we use the entanglement-based theoretical description of SR-CV-QKD to analyze the secret key rate under individual and collective Gaussian attacks. Section~\ref{sec:experiment} presents the details of the experimental demonstration of the protocol's feasibility. Finally, our conclusions are summarized in Sec.~\ref{sec:conc}.

\section{Details of the SR-CV-QKD protocol} 
\label{sec:SR-CV-QKD}

In this section, we describe the prepare-and-measure version of the protocol, which corresponds to the actual physical implementation. The equivalent entanglement-based description is presented in Sec.~\ref{sec:theory}.

In each round of SR-CV-QKD, Alice chooses two independent Gaussian random variables $(q_A, p_A)$, both distributed as $\mathcal{N}(0,V_A)$, and sends Bob the coherent state $\ket{q_A + ip_A}$, which we refer to as the \emph{signal pulse}. In addition, she sends a coherent-state \emph{reference pulse} in the next time bin. The mean quadrature values of the reference pulse in Alice's reference frame, ($q_{A_R}$, $p_{A_R}$),\footnote{We use lowercase letters $q$ and $p$ for c-number mean values of quadratures and capital letters $Q$ and $P$ for quadrature operators.} are publicly known. The amplitude of the reference pulse,\footnote{Throughout this paper, we express all quadrature variances and correlations in shot-noise units, $N_0$, and all quadrature values and amplitudes in units of $N_0^{1/2}$.} $V_R^{1/2} = (q_{A_R}^2 + p_{A_R}^2)^{1/2}$, is fixed and may be several times larger than $V_A^{1/2}$, but much smaller than that of a typical LO. Using reference pulses with a relatively small amplitude is a practically important aspect of SR-CV-QKD, which helps to reduce the interference with the signal pulse, as compared to the effect of a large-amplitude (classical) pulse, whose ``long tail'' cannot be completely suppressed and hence would interfere with the signal if time multiplexed at the same rate.

In each round, Bob performs a homodyne measurement of one of the quadratures ($Q_B$ or $P_B$) of the received signal pulse to estimate its mean value ($q_B$ or $p_B$, respectively), where these quadratures are defined relative to his own high-power LO. He also performs a heterodyne measurement on the received reference pulse to obtain both of its mean quadrature values, $q_{B_R}$ and $p_{B_R}$ (again, with respect to his LO).

\begin{figure}[t]
	\centering
	\includegraphics[width=1.0\linewidth]{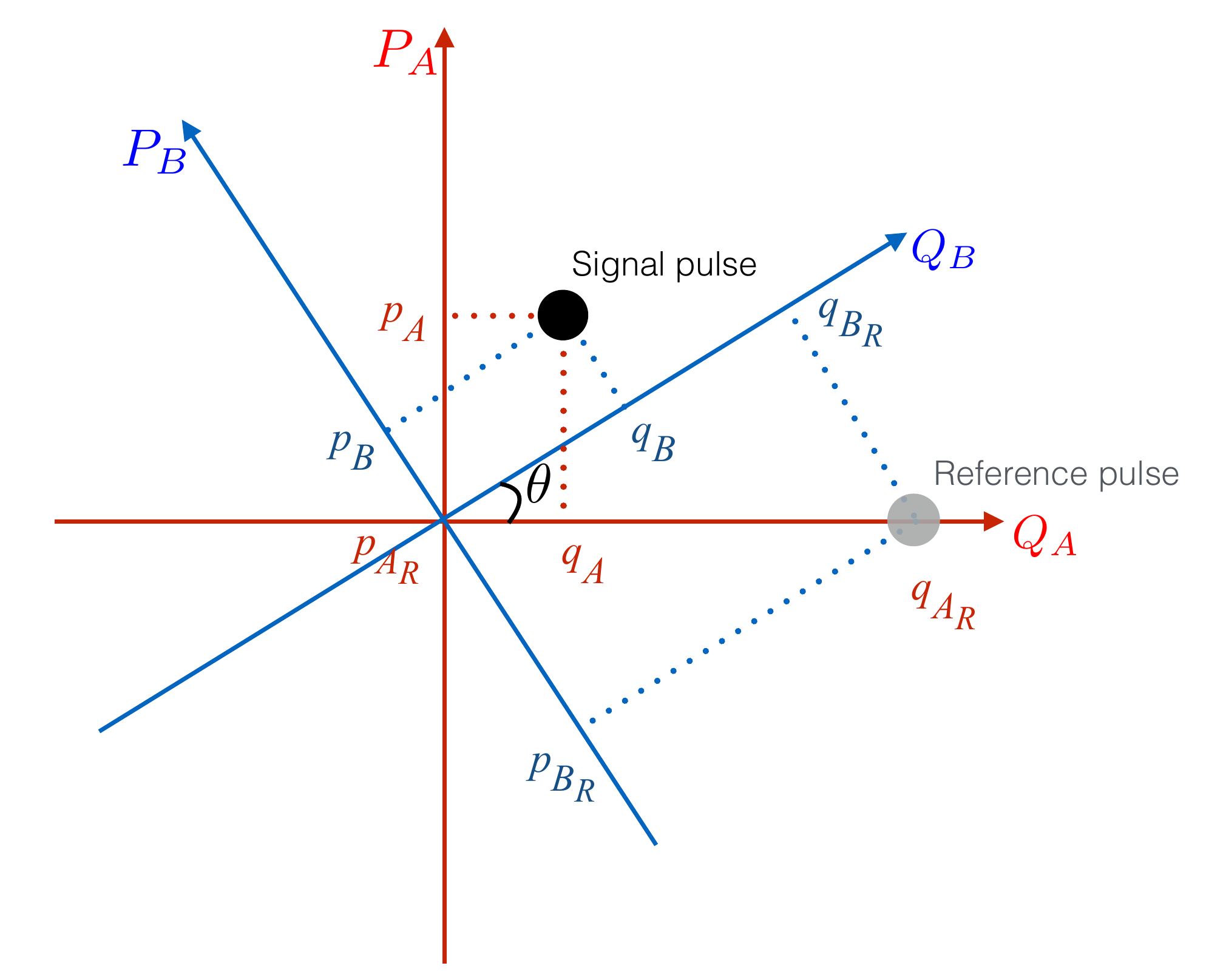}
	\caption{The phase-space representation of Alice's and Bob's misaligned reference frames. The reference pulse is assumed to be on the $Q_A$ axis for Alice in this example, and the signal pulse is randomly placed.}
	\label{fig:phasespace}
\end{figure}

The phase-space representation of Alice's and Bob's misaligned reference frames is shown in Fig.~\ref{fig:phasespace}. The phase difference $\theta$ between Alice's and Bob's frames is a time-dependent quantity since their individual LOs are free-running, and we assume that $\theta$ at any time is a random variable distributed uniformly on $(-\pi,\pi]$ and that the frequency of its fluctuations (i.e., the phase noise bandwidth), $f_{\theta}$, which is measured and calibrated before the protocol begins, is much lower than the rate of pulse generation. In other words, let $\Delta t$ be the time delay between signal and reference plus the duration of both pulses. Then, we require that this duration is much shorter than the inverse of the bandwidth $f_{\theta}$, i.e.,
\begin{equation}
	\Delta t \ll f_{\theta}^{-1}.
	\label{eq:delta_t}
\end{equation} 
Provided that condition~\eqref{eq:delta_t} is satisfied, the $\theta$ value will be the same for measurements on both pulses. We note that \erf{eq:delta_t} also places a restriction on the phase stability of Alice's laser source; specifically, this source should be phase stable over the time period specified by $\Delta t$. The same phase stability is required of Bob's LO. 

Estimation of the phase difference $\theta$ is the key element in SR-CV-QKD. Since Bob knows the mean quadrature values of the reference pulse both in Alice's frame, ($q_{A_R}$, $p_{A_R}$), and in his own frame, ($q_{B_R}$, $p_{B_R}$), he can calculate an estimate $\hat\theta$ of the phase difference, via:
\beq
\label{eq:thetahat_matrix}
\begin{pmatrix}
 q_{B_R} \\ p_{B_R} 
\end{pmatrix}
= \sqrt{T_{\mathrm{eff}}}
\begin{pmatrix}
 \cos \hat\theta & -\sin \hat\theta \\
 \sin \hat\theta & \cos \hat\theta 
\end{pmatrix}
\begin{pmatrix}
 q_{A_R} \\ p_{A_R} 
\end{pmatrix},
\eeq
where $0 < T_{\mathrm{eff}} \leq 1$ is the effective channel transmittance that can be eliminated to obtain
\beq
\label{eq:thetahat_ratio}
\hat\theta = \tan^{-1} \left( \frac{p_{B_R} q_{A_R} - q_{B_R} p_{A_R}}{q_{B_R} q_{A_R} + p_{B_R} p_{A_R}} \right).
\eeq
In the following we will assume without loss of generality that Alice's reference pulse has $p_{A_R} = 0$, in which case Eq.~\eqref{eq:thetahat_ratio} becomes
\begin{equation}
\label{eq:thetahat_ratio_1}
\hat\theta = \tan^{-1} \left( \frac{p_{B_R}}{q_{B_R}} \right) .
\end{equation}
Since the reference pulse has a relatively small amplitude, its quantum uncertainty cannot be ignored, and therefore even in the case of a technically ideal measurement, there will be an error in the phase difference estimate, i.e.,
\beq
\label{eq:thetas}
\hat\theta = \theta + \varphi, 
\eeq
where the estimation error $\varphi$ is a random variable distributed according to some probability distribution $\mathcal{P}(\varphi)$. We assume that $\theta$ and $\varphi$ are independent random variables, since they arise from separate physical processes. We will see in Sec.~\ref{sec:theory} that the error of phase difference estimation plays a critical role in determining the expected secret key rate of SR-CV-QKD. 

As is standard in modern CV-QKD, our protocol employs reverse reconciliation. Bob sends Alice his estimate of the phase difference between their frames, $\hat\theta$, and which quadrature of the signal pulse ($Q_B$ or $P_B$) he measured. Then Alice rotates her tabulated values for the signal pulse by $\hat\theta$, to obtain an estimate ($\hat{q}_B$ or $\hat{p}_B$) of Bob's measured quadrature value, via
\beq
\begin{pmatrix}
 \hat{q}_B \\ \hat{p}_B 
\end{pmatrix}
= \sqrt{T_{\mathrm{eff}}}
\begin{pmatrix}
 \cos \hat\theta & -\sin \hat\theta \\
 \sin \hat\theta & \cos \hat\theta 
\end{pmatrix}
\begin{pmatrix}
 q_A \\ p_A 
\end{pmatrix}. 
\eeq
At this point Alice and Bob share a partially correlated Gaussian random variable and the remainder of the protocol is the same as conventional CV-QKD, which proceeds by performing channel estimation, error correction, and privacy amplification after a large enough block of (imperfect) shared randomness has been collected \cite{Grosshans2003, jouguet2013experimental}.

We mention a number of points about the new protocol before examining it quantitatively. First, the occurrence ratio of reference pulses to signal pulses does not have to be 1:1. If the phase drift is significantly slower than the signal pulse rate, then one can utilize fewer reference pulses to estimate the slow drifting phase difference. In this situation, condition~\eqref{eq:delta_t} should hold for $\Delta t$ being the period between subsequent reference pulses.

Second, the frequency difference between Alice's source and Bob's LO should be reasonably stable: this can be accomplished through the use of single-frequency lasers locked individually to a stable reference frequency such as an atomic line. In case the stable frequency reference is difficult to implement, commercially available single-frequency lasers can still be utilized. If the relative frequency drift is slow and the linewidth is sufficiently narrow, the drift can be treated as a phase noise, which the SR-CV-QKD scheme will handle. However, if the relative frequency drift is fast, one should consider frequency locking Bob's LO by utilizing a dedicated locking beam and utilizing the \emph{transfer cavity} technique \cite{riedle1994stabilization, bohlouli2006optical}. We note that synchronizing separate laser sources may turn out to be a particularly challenging task in circumstances where practical limitations on laser properties exist (e.g., limited coherence properties of available on-chip lasers for integrated photonics implementations). 

Third, utilizing a heterodyne detector and a homodyne detector in concert (heterodyne for reference pulses and homodyne for signal pulses) can be challenging since consecutive pulses need to be routed to the correct detector. This can be accomplished by either actually having two types of detectors and routing each pulse accordingly, or, more practically, by frequency detuning the local oscillator from the carrier frequency when a heterodyne detection is required. Such small detuning can be performed at GHz rates and thus this solution is valid if the pulses are nanosecond separated. However, in some instances Bob may be restricted to performing homodyne measurements only, for example, if it is desirable to simplify his receiver hardware as much as possible. In this case Alice can send a pair of closely spaced twin reference pulses, and Bob will perform orthogonal quadrature measurements on them sequentially, obtaining $q_{B_R}$ from the measurement on one reference pulse and $p_{B_R}$ from the measurement on the other. In addition to hardware simplification, this \emph{twin-reference-pulse} mode results in a lower uncertainty of the phase difference estimate, as quantified in Sec.~\ref{sec:theory} below, but at the expense of a reduction in the number of time bins available for signal pulses (which constitute the raw data for the eventual key) for a fixed communication time.

Finally, we note that using reference pulses to perform phase drift estimation is not only useful for the SR-CV-QKD protocol \emph{per se}, but also for calibration purposes before and during the protocol. This technical improvement is discussed in more detail in Appendix~\ref{sec:calibration}.

\section{Secret key rate analysis}
\label{sec:theory}

While claims of secure key distribution should be based on empirically estimated correlations between Alice and Bob \cite{Leverrier:2015tr}, it is common to calculate an expected secret key rate based on reasonable assumptions on the communication channel and detection apparatus. The usefulness of such a theoretical analysis is in revealing the effects of various design parameters on the achievable key rate. This calculation is particularly important in our case since it allows us to compare the expected performance of SR-CV-QKD against the respective conventional protocol that requires LO transmission.

We follow the approach in Refs.~\cite{Scarani:2009wq, GarciaPatron:2007ws} and compute the asymptotic expected key rate in the presence of a lossy, noisy passive Gaussian process, $\mathcal{E}$, that models channel transmittance, channel excess noise, detection inefficiency, and electronic detector noise. The entanglement-based description of the conventional protocol begins with the density matrix for the state shared between Alice and Bob before they perform any measurements: $\rho_{AB} = \mathcal{E}(\rho_{SV})$, where $\rho_{SV}$ is the ideal two-mode squeezed vacuum state. Since $\rho_{SV}$ and $\mathcal{E}$ are Gaussian, one can equivalently express the state $\rho_{AB}$ in terms of its covariance matrix (represented in the basis $\{Q_A, P_A, Q_B, P_B\}$)~\cite{Scarani:2009wq, GarciaPatron:2007ws}:
\begin{equation}
\label{eq:gamma}
\gamma_{AB} = \begin{pmatrix} V \openone & C \sigma_z \\ 
C \sigma_z & T \eta (V + \chi) \openone \end{pmatrix} 
\end{equation}
with 
\begin{equation}
\label{eq:C}
C = \sqrt{T \eta (V^2 - 1)} , 
\end{equation}
where $\openone = {\scriptstyle \begin{pmatrix} 1 & 0 \\ 0 & 1 \end{pmatrix}}$ and $\sigma_z = {\scriptstyle \begin{pmatrix} 1 & 0 \\ 0 & -1 \end{pmatrix}}$. In Eqs.~\eqref{eq:gamma} and \eqref{eq:C}, $T$ is the channel transmittance, $\eta$ is the detector efficiency (so the overall effective transmittance is $T_{\mathrm{eff}} = T \eta$), $\chi$ is the channel noise (referred to the input of the channel), and $V$ is the variance of both quadratures of Alice's output state, i.e., $V = V_A + 1$, where $V_A$ is the variance of Alice's Gaussian modulation of the signal pulse. The noise can be modeled as a sum of three terms~\cite{Scarani:2009wq}:
\begin{equation}
\label{eq:chi}
\chi = \frac{1 - T \eta}{T \eta} + \frac{V_{\mathrm{el}}}{T \eta} + \varepsilon ,
\end{equation}
where the first term is the loss-induced vacuum noise, the second term is the contribution of the detector electronic noise with the variance $V_{\mathrm{el}}$, and $\varepsilon$ is the excess noise in the channel. Note that this noise model treats channel and detector contributions on equal footing, thus resulting in a conservative estimate of the expected key rate. Some works~\cite{Grosshans2003, lodewyck2007quantum, Fossier.JPB.2009, Qi:2015uf} use a more nuanced model which assumes that Eve cannot benefit from the noise added by Bob's detector, therefore resulting in a more optimistic key rate estimate. In this work, we use the conservative noise model of Ref.~\cite{Scarani:2009wq}, which corresponds to a stronger security scenario.

In SR-CV-QKD, in addition to the process $\mathcal{E}$, we need to take into account the effect of phase-space rotations due to the reference frame mismatch, including averaging over distributions of random variables $\theta$ and $\varphi$. The resulting density matrix for the state shared between Alice and Bob before they perform any measurements is
\beq
\overline\rho_{AB} 
= \overline{\rho_{AB}(\hat\theta,\theta)}
= \int_{-\pi}^{\pi} d\varphi \mathcal{P}(\varphi) \int_{-\pi}^{\pi} \frac{d\theta}{2\pi} 
\rho_{AB}(\hat\theta,\theta)
 \label{eq:rho_ab_bar}
\eeq
with
\beq
\rho_{AB}(\hat\theta,\theta) = U_A(-\hat\theta) U_B(\theta) \rho_{AB} U\dg_A(-\hat\theta) U\dg_B(\theta) ,
\eeq
where $U_{A(B)}(\phi)$ is the operator of a phase-space rotation of Alice's (Bob's) mode by angle $\phi$. The interpretation of this state is that Bob's mode undergoes a rotation by the angle equal to the actual phase difference $\theta$, and Alice attempts to compensate for this by applying a rotation of her mode by the angle $-\hat\theta$. Note that in the ideal case where $\hat\theta = \theta$, these rotations describe an orthogonal transformation of Bob's state and the conjugate orthogonal transformation \cite{Leverrier:2009hu} of Alice's state, whose combination leaves the bipartite state invariant, i.e., $\rho_{AB}(\theta,\theta) = \rho_{AB}$ and, consequently, $\overline\rho_{AB} = \rho_{AB}$ in the ideal case. Thus, in the entanglement-based description, the SR-CV-QKD protocol can be seen as an attempt to restore the imperfect EPR correlations in the state $U_B(\theta) \rho_{AB} U\dg_B(\theta)$ by compensating for the random rotation experienced by Bob's mode. 

The state $\overline\rho_{AB}$ is also Gaussian, and its covariance matrix can be expressed as
\beq
\overline\gamma_{AB}
= \overline{\gamma_{AB}(\hat\theta,\theta)}
= \int_{-\pi}^{\pi} d\varphi \mathcal{P}(\varphi) \int_{-\pi}^{\pi} \frac{d\theta}{2\pi} 
\gamma_{AB}(\hat\theta,\theta)
 \label{eq:Gamma_ab_bar}
\eeq
with
\beq
\gamma_{AB}(\hat\theta,\theta) = \left[ \mathsf{U}_A(-\hat\theta) \oplus \mathsf{U}_B(\theta) \right] 
\gamma_{AB} \left[ \mathsf{U}^{\mathsf{T}}_A(-\hat\theta) \oplus \mathsf{U}^{\mathsf{T}}_B(\theta) \right], 
\eeq
where $\mathsf{U}_{A(B)}$ is the symplectic representation of a phase-space rotation operator $U_{A(B)}$~\cite{Weedbrook:2012tz}. Computing these rotations and integrals yields
\begin{equation}
\label{eq:gamma_bar}
\overline\gamma_{AB} = \begin{pmatrix} V \openone & C \overline\Phi \\ 
C \overline\Phi & T \eta (V + \chi) \openone \end{pmatrix} 
\end{equation}
with
\begin{eqnarray}
& \overline\Phi = \begin{pmatrix} \overline{\cos\varphi} & \overline{\sin\varphi} \\
\overline{\sin\varphi} & - \overline{\cos\varphi} \end{pmatrix} , & \\
& \begin{array}{c} 
\overline{\cos\varphi} = \int_{-\pi}^{\pi} d\varphi \mathcal{P}(\varphi) \cos\varphi , \\
\overline{\sin\varphi} = \int_{-\pi}^{\pi} d\varphi \mathcal{P}(\varphi) \sin\varphi . \end{array} & 
\end{eqnarray}
Comparing the covariance matrix $\overline\gamma_{AB}$ of Eq.~\eqref{eq:gamma_bar} to $\gamma_{AB}$ of Eq.~\eqref{eq:gamma}, we see that the effect of the reference frame alignment in SR-CV-QKD is to replace $\sigma_z$ by $\overline\Phi$ in off-diagonal blocks. In the following we will assume that the phase estimation error is dominated by the quantum uncertainty of the reference pulse(s), in which case the distribution $\mathcal{P}(\varphi)$ is symmetric around $\varphi = 0$, and consequently $\overline{\sin\varphi} = 0$. Then, 
\begin{equation}
\overline\Phi = \overline{\cos\varphi}\, \sigma_z, 
\end{equation}
and the effect of the reference frame alignment is to simply rescale the $\expect{Q_A Q_B}$ and $\expect{P_A P_B}$ correlations by the factor $\overline{\cos\varphi}$. Therefore we can carry over the analysis in Refs.~\cite{Scarani:2009wq, GarciaPatron:2007ws} for the respective conventional CV-QKD protocol, and simply replace the off-diagonal blocks in the covariance matrix by the scaled versions.

A more visual way for evaluating variances and correlations in SR-CV-QKD is by using the Heisenberg picture, i.e., applying the phase-space rotations to the quadrature operators,
\begin{subequations}
\begin{align}
\begin{pmatrix} Q_A(\hat\theta) \\ P_A(\hat\theta) \end{pmatrix}
= \begin{pmatrix}
 \cos \hat\theta & -\sin \hat\theta \\
 \sin \hat\theta & \cos \hat\theta 
\end{pmatrix}
\begin{pmatrix} Q_A \\ P_A \end{pmatrix}, \\
\begin{pmatrix} Q_B(\theta) \\ P_B(\theta) \end{pmatrix}
= \begin{pmatrix}
 \cos \theta & \sin \theta \\
 -\sin \theta & \cos \theta 
\end{pmatrix}
\begin{pmatrix} Q_B \\ P_B \end{pmatrix},
\end{align}
\end{subequations}
and evaluating quantum expectation values over the unrotated state $\rho_{AB}$, as well as averaging over distributions of random variables $\theta$ and $\varphi$. Using elements of the covariance matrix $\gamma_{AB}$, it is straightforward obtain:
\begin{subequations}
\label{eq:Q-vars}
\begin{align}
& \langle Q_A^2 \rangle = \int_{-\pi}^{\pi} d\varphi \mathcal{P}(\varphi) 
\int_{-\pi}^{\pi} \frac{d \theta}{2\pi} 
\mathrm{Tr}\left[ \rho_{AB} Q_A^2 (\hat\theta) \right]
= V , \label{eq:EV_XA2} \\
& \langle Q_B^2 \rangle = \int_{-\pi}^{\pi} \frac{d \theta}{2\pi} 
\mathrm{Tr}\left[ \rho_{AB} Q_B^2 (\theta) \right]
= T \eta (V + \chi) , \label{eq:EV_XB2}
\end{align}
\begin{align}
\langle Q_A Q_B \rangle & = \int_{-\pi}^{\pi} d\varphi \mathcal{P}(\varphi) 
\int_{-\pi}^{\pi} \frac{d \theta}{2\pi} 
\mathrm{Tr}\left[ \rho_{AB} Q_A (\hat\theta) Q_B (\theta) \right] \nonumber \\
& = C \overline{\cos\varphi} , \label{eq:EV_XAXB}
\end{align}
\end{subequations}
and analogously for other variances and correlations, thus reproducing the elements of $\overline\gamma_{AB}$ in Eq.~\eqref{eq:gamma_bar}.

Alice's preparation of Gaussian-modulated coherent states in the prepare-and-measure description corresponds to her performing a heterodyne measurement on the state of mode $A$ in the entanglement-based description. This heterodyne measurement is equivalent to mixing mode $A$ with vacuum on a balanced beam splitter and performing homodyne measurements on conjugate quadratures of two output modes $A'$ and $A''$. Due to the symmetry between the two quadratures, it is sufficient to consider the measurement of the $Q$ quadrature of mode $A'$, which is given by
\begin{equation}
Q_{A'} = \frac{1}{\sqrt{2}} \left( Q_A + Q_{\mathrm{vac}} \right).
\end{equation}
Since the vacuum noise is not correlated with any other mode, it is easy to obtain:
\begin{subequations}
\label{eq:Q-vars-het}
\begin{align}
& \langle Q_{A'}^2 \rangle = \frac{1}{2} \left(\langle Q_A^2 \rangle + 1 \right) = \frac{1}{2} (V + 1) , \\
& \langle Q_{A'} Q_B \rangle = \frac{1}{\sqrt{2}} \langle Q_A Q_B \rangle 
= \frac{1}{\sqrt{2}} C \overline{\cos\varphi} .
\end{align}
\end{subequations}

\subsection{Individual attacks}

The asymptotic secret key rate against individual attacks for reverse reconciliation is given by
\beq
K_{\mathrm{ind}} = \beta I_{A'B} - I_{EB},
\eeq
where $0 < \beta \leq 1$ is the reconciliation efficiency, $I_{A'B}$ is the mutual information between Alice's and Bob's measurements, and $I_{BE}$ is the mutual information between Eve's and Bob's measurements. These mutual informations are given by
\begin{subequations}
\begin{align}
I_{A' B} & = \frac{1}{2} \log_2 \left( \frac{V_{B}}{V_{B | A'}} \right) , \label{eq:Iab}\\
I_{EB} & = \frac{1}{2} \log_2 \left( \frac{V_B}{V_{B|E}}\right), \label{eq:Ibe}
\end{align}
\end{subequations}
where $V_{B} = \langle Q_{B}^2 \rangle$ is the variance of the quadrature $Q_{B}$ measured by Bob, and 
\begin{equation}
V_{B | A'} = \langle Q_{B}^2 \rangle - \frac{\langle Q_{A'} Q_{B} \rangle^2}{\langle Q_{A'}^2 \rangle}
\end{equation}
is the conditional variance that quantifies Alice's uncertainty on $Q_{B}$ after the measurement of $Q_{A'}$. Using Eqs.~\eqref{eq:Q-vars} and \eqref{eq:Q-vars-het}, we obtain 
\begin{equation}
\label{eq:V_BAp}
V_{B | A'} = T \eta [ \chi + 1 + (V-1) \xi ]
\end{equation}
and
\begin{equation}
I_{A' B} = \frac{1}{2} \log_2 \left[ 
\frac{V + \chi}{\chi + 1 + (V-1) \xi} \right] ,
\end{equation}
where 
\begin{equation}
\xi = 1 - (\overline{\cos\varphi})^2 .
\end{equation}

We see from Eq.~\eqref{eq:V_BAp} that the effect of using the reference pulse is the increase in the conditional variance by the additional term 
\begin{equation}
\label{eq:V_add}
\Delta V_{B | A'} = V_{B | A'} - \left[ V_{B | A'} \right]_{\xi = 0} 
=  T \eta (V-1) \xi .
\end{equation}
If the distribution $\mathcal{P}(\varphi)$ is tight, then $\overline{\cos\varphi} \approx 1 - \frac{1}{2} \overline{\varphi^2}$ and $\xi \approx \overline{\varphi^2}$. Also, if condition~\eqref{eq:delta_t} is satisfied, the $\theta$ value is constant during each $\hat\theta$ estimation, and the variance of the estimated value is
$V_{\hat\theta} = V_{\varphi} = \overline{\varphi^2}$ (recall that we assume symmetric $\mathcal{P}(\varphi)$, which implies $\overline{\varphi} = 0$). If $\mathcal{P}(\varphi)$ monotonically and rapidly decreases with $|\varphi|$ from the maximum value at $\varphi = 0$, the variance $V_{\hat\theta} = \overline{\varphi^2}$ is a tight upper bound on $\xi$, i.e., $\xi \lessapprox V_{\hat\theta}$, and, consequently,
\begin{equation}
\label{eq:V_add_bound_1}
\Delta V_{B | A'} \lessapprox  T \eta (V-1) V_{\hat\theta}. 
\end{equation}
We can evaluate $V_{\hat\theta}$ by expressing it as
\begin{equation}
\label{eq:V_theta_1}
V_{\hat\theta} = \int_{-\pi}^{\pi} \frac{d \theta}{2\pi}
\left[ \left(\frac{\partial \hat\theta}{\partial z} \right)^2 V_z \right]_{\hat\theta = \theta} ,
\end{equation}
where $z = \tan \hat\theta = p_{B_R} / q_{B_R}$, $\partial \hat\theta / \partial z = \cos^2 \hat\theta$, and 
\begin{equation}
\label{eq:V_z}
V_z = \left( \frac{V_{Q_{B_R}} + \delta_R}{q_{B_R}^2} + \frac{V_{P_{B_R}} + \delta_R}{p_{B_R}^2} \right) z^2
\end{equation}
is the dimensionless variance of the measured value of $z$. In Eq.~\eqref{eq:V_z}, $\delta_R = 1$ in the single-reference-pulse mode (a heterodyne measurement is performed on a single reference pulse) and $\delta_R = 0$ in the twin-reference-pulse mode (sequential homodyne measurements are performed on a pair of twin reference pulses). In general, the value of $V_z$ depends on the modulation used by Alice to generate reference pulses. In particular, for fixed mean quadrature values $q_{A_R} = V_R^{1/2}$ and $p_{A_R} = 0$, we find: 
\begin{subequations}
\begin{align}
& V_{Q_{B_R}} = V_{P_{B_R}} = T \eta (\chi + 1) , \\
& q_{B_R}^2 = T \eta V_R \cos^2 \hat\theta, \\
& p_{B_R}^2 = T \eta V_R \sin^2 \hat\theta. 
\end{align}
\end{subequations}
By substituting these expressions into Eqs.~\eqref{eq:V_z} and \eqref{eq:V_theta_1}, we obtain:
\begin{equation}
\label{eq:V_theta_2}
V_{\hat\theta} = \frac{\chi + 1}{V_R} + \frac{\delta_R}{T \eta V_R}.
\end{equation}
Correspondingly, the tight upper bound on the conditional variance increase due to the reference pulse use, given by Eq.~\eqref{eq:V_add_bound_1}, can be now expressed in terms of experimental parameters:
\begin{equation}
\label{eq:V_add_bound_2}
\Delta V_{B' | A'} \lessapprox \frac{(V-1) [T \eta (\chi + 1) + \delta_R]}{ V_R},
\end{equation}
which scales as $V_A / V_R$. The corresponding lower bound on the mutual information between Alice and Bob is
\begin{equation}
I_{A' B} \gtrapprox \frac{1}{2} \log_2 \left[ 
\frac{V + \chi}{(\chi + 1)\left(1+ \frac{V-1}{V_R}\right) + \frac{(V-1) \delta_R}{T\eta V_R} } \right] ,
\label{eq:Iab_bound}
\end{equation}

Now, to evaluate the mutual information between Eve and Bob, one can apply the Heisenberg uncertainty relation to the pure state held by Bob conditioned on Alice's and Eve's measurements, to obtain \cite{Scarani:2009wq}:
\begin{equation}
\label{eq:HUR}
V_{B|E} V_{B|A} \geq 1 ,
\end{equation}
where, due to the symmetry between $Q$ and $P$ quadratures, $B$ stands for any quadrature of Bob's mode. By substituting inequality~\eqref{eq:HUR} into \erf{eq:Ibe}, we obtain: 
\begin{align}
I_{EB} \leq & \frac{1}{2}\log_2 \left( V_B V_{B|A} \right) \nn \\
= & \frac{1}{2}\log_2 \left\{ \frac{(T\eta)^2 (V+\chi)}{V} \left[V\chi + 1 + (V^2-1)\xi\right] \right\} \nn \\
\lessapprox & \frac{1}{2}\log_2 \left\{ \frac{(T\eta)^2 (V+\chi)}{V} \bigg[ V\chi + 1 \right. \nn \\
& \left. + (V^2-1)\left( \frac{\chi + 1}{V_R} + \frac{\delta_R}{T \eta V_R} \right) \bigg] \right\},
\label{eq:Ibe_bound}
\end{align}
where in the last line we have used $\xi \lessapprox V_{\hat\theta}$ to express the bound on mutual information in terms of experimental parameters. 

Finally, putting together the bounds in Eqs.~\eqref{eq:Iab_bound} and \eqref{eq:Ibe_bound}, we obtain the minimum key rate that is secure against individual attacks, $K_{\mathrm{ind}} \geq K_{\mathrm{ind}}^{\mathrm{min}}$, expressed in terms of experimental parameters:
\begin{align}
K_{\mathrm{ind}}^{\mathrm{min}} = & 
\frac{\beta}{2} \log_2 \left[ \frac{V + \chi}{(\chi + 1)
\left(1 + \frac{V-1}{V_R}\right) + \frac{(V-1) \delta_R}{T\eta V_R} } \right] \nn \\
& - \frac{1}{2}\log_2 \left\{ \frac{(T\eta)^2 (V+\chi)}{V} \bigg[ V\chi + 1 \right. \nn \\
& \left. + (V^2-1)\left( \frac{\chi + 1}{V_R} + \frac{\delta_R}{T \eta V_R} \right) \bigg] \right\}.
\label{eq:shannon_rate}
\end{align}
Note that all terms associated with the reference pulse's quantum uncertainty reduce the key rate and scale inversely with $V_R$. In the limit of a large-amplitude (classical) reference pulse, $V_R \rightarrow \infty$, the theoretical key rate of SR-CV-QKD is the same as that for the respective conventional CV-QKD protocol that requires LO transmission~\cite{Scarani:2009wq, GarciaPatron:2007ws}. However, even in this limit, SR-CV-QKD could still be practically advantageous since it avoids many technical difficulties associated with LO transmission, as detailed in Sec.~\ref{sec:intro}.

\subsection{Collective attacks}

The expected secret key rate against collective attacks for reverse reconciliation is given by  
\beq
K_{\mathrm{col}} = \beta I_{A'B} - \chi_{BE},
\eeq
where the lower bound on $I_{A'B}$ is given by Eq.~\eqref{eq:Iab_bound} and $\chi_{BE}$ is the Holevo quantity for Eve's maximum accessible information. For Gaussian protocols, the Holevo quantity is \cite{lodewyck2007quantum, GarciaPatron:2007ws}
\beq
\chi_{BE} = S(\overline\rho_{AB}) - S(\overline\rho_A^{q_B}),
\eeq
where $S(\rho)$ is the von Neumann entropy of the state $\rho$, $\overline\rho_{AB}$ is the state shared between Alice and Bob before they perform any measurements, and $\overline\rho_A^{q_B}$ is the state of Alice's system conditional on Bob's measurement outcome $q_B$. Since these states are Gaussian, the entropy is evaluated in terms of symplectic eigenvalues of the covariance matrices of the corresponding states \cite{Weedbrook:2012tz}. This procedure results in \cite{GarciaPatron:2007ws}
\beq
\chi_{BE} = G\left(\frac{\lambda_1-1}{2}\right)+G\left(\frac{\lambda_2-1}{2}\right)-G\left(\frac{\lambda_3-1}{2}\right),
\eeq
where $G(x) = (x+1)\log_2(x+1) - x\log_2(x)$, and the eigenvalues $\lambda_i$ are obtained from
\begin{subequations}
\label{eq:lambda_all}
\begin{eqnarray}
\lambda_{1,2}^2 & = & \frac{1}{2}(\Delta \pm \sqrt{\Delta^2 - 4 D^2}), \label{eq:lambda12} \\
\Delta & = & V^2 (1 - 2 T \eta) + (T\eta)^2 (V+\chi)^2 \nn \\
&& + 2 T \eta \left[ 1 + (V^2-1)\xi \right], \\
D & = & T \eta \left[ V\chi+1 + (V^2-1)\xi \right], \\
\lambda_3^2 & = & V \frac{V\chi + 1 + (V^2-1)\xi}{V+\chi}. \label{eq:lambda3}
\end{eqnarray}
\end{subequations}
Furthermore, we find that $\chi_{BE}$ monotonically increases as $\xi$ increases, and therefore we can upper bound $\chi_{BE}$ using $\xi \leq V_{\hat\theta}$. Thus, replacing $\xi$ in Eqs.~\eqref{eq:lambda_all} with expression~\eqref{eq:V_theta_2} for $V_{\hat\theta}$ completes the derivation of the minimum expected key rate that is secure against collective attacks, $K_{\mathrm{col}} \geq K_{\mathrm{col}}^{\mathrm{min}}$, in terms of experimental parameters. We note that since SR-CV-QKD is a Gaussian protocol, security against collective attacks is sufficient for asymptotic unconditional security (i.e., security against coherent attacks) with some processing overhead \cite{Renner:2009jg, Leverrier:2013jq}.

\begin{figure}[!tb]
\centering
\includegraphics[width=1.0\linewidth]{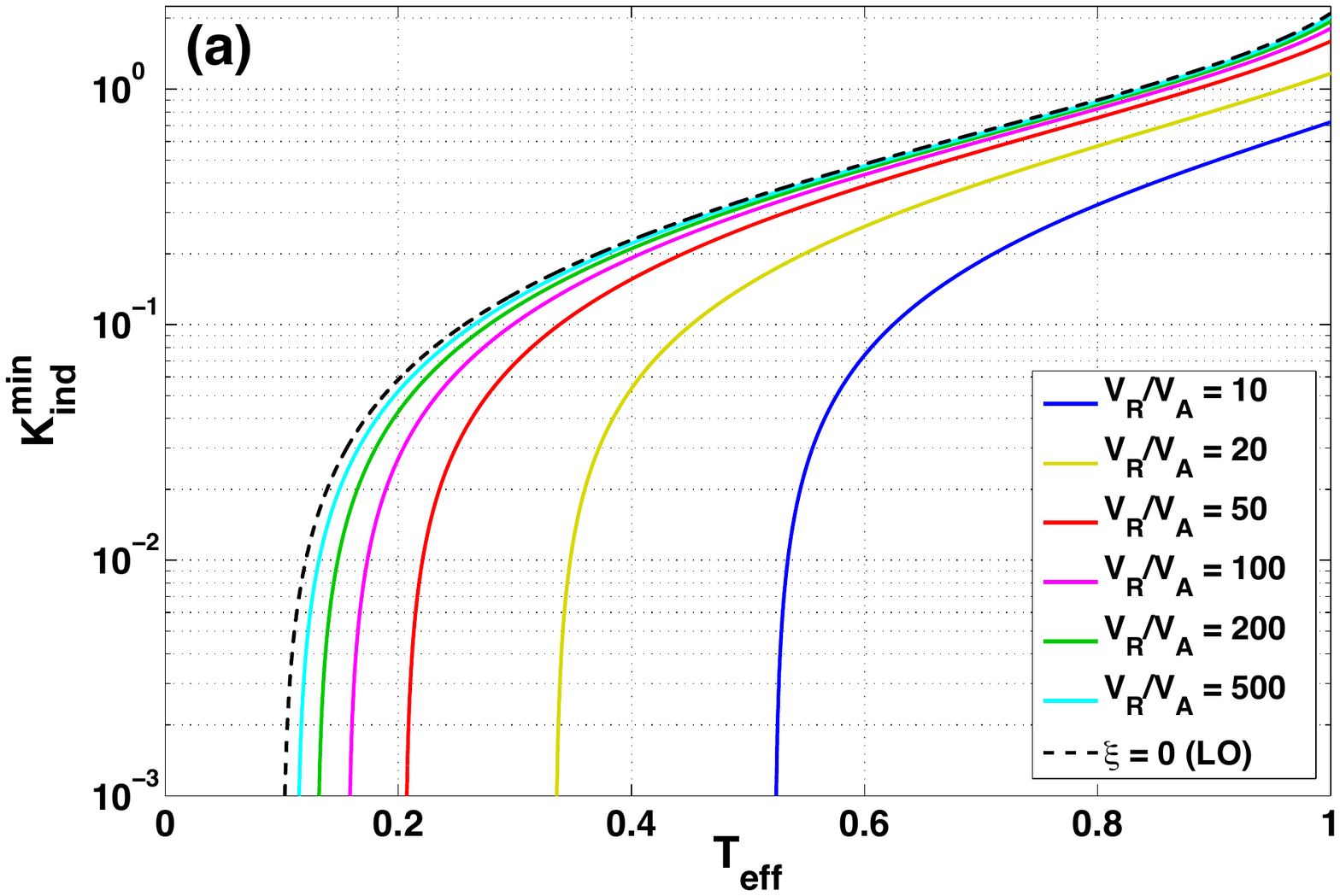}
\includegraphics[width=1.0\linewidth]{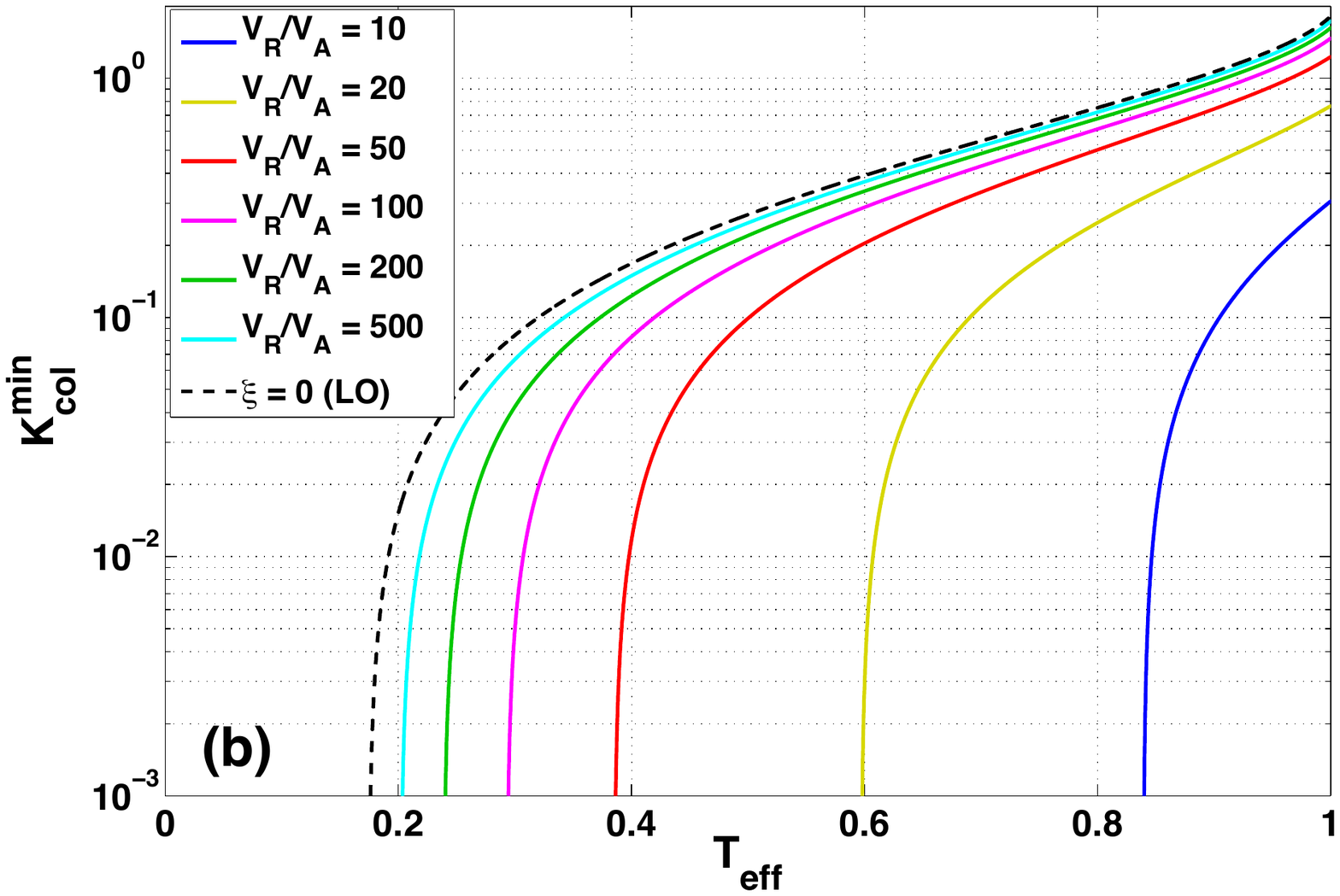}
\caption{Expected minimum key rates for the SR-CV-QKD protocol, secure against (a) individual attacks and (b) collective attacks, as functions of the effective transmittance $T_{\mathrm{eff}} = T \eta$. The parameter values used are $V_A = 40$, $\varepsilon = 0.01$, $V_{\rm el} = 0.01$ (all in shot-noise units), $\beta=0.95$, $\delta_R = 1$. As shown in the legend, different curves correspond to different values of the reference-pulse amplitude $V_R^{1/2}$ (specifically, $V_R/V_A = \{10,20,50,100,200,500\}$), along with the curve for $\xi = 0$, in which case the key rate is the same as that for the respective conventional CV-QKD protocol with LO transmission. The curves terminate on the left at values of $T_{\mathrm{eff}}$ below which the expected secret key rate is zero.}
\label{fig:keyrates}
\end{figure}

Figure \ref{fig:keyrates} shows the expected minimum key rates secure against individual and collective attacks, as expressed by $K_{\mathrm{ind}}^{\mathrm{min}}$ and $K_{\mathrm{col}}^{\mathrm{min}}$, respectively. Each plot shows the key rate as a function of the effective transmittance $T_{\mathrm{eff}} = T \eta$, for a number of reference-pulse amplitude values ($V_R/V_A = \{10,20,50,100,200,500\}$ with $V_A = 40$) and also for $\xi = 0$ (which results in the same key rate as the respective conventional CV-QKD protocol with LO transmission). We see that as $V_R$ increases the performance of SR-CV-QKD approaches that of conventional CV-QKD, and at $V_R \sim 500 V_A$ the achievable key rates are very similar for the two protocols. Figure \ref{fig:keyrates} shows results for the single-reference-pulse mode ($\delta_R = 1$); the results for the twin-reference-pulse mode ($\delta_R = 0$) are very similar, except that smaller values of $V_R/V_A$ are required to approach the $\xi = 0$ curve, which is a result of the higher accuracy of the phase difference estimation possible in this mode of operation.

\section{Experimental characterization and demonstration} 
\label{sec:experiment}

\begin{figure}[!tb]
\centering
{\includegraphics[width=0.95\linewidth]{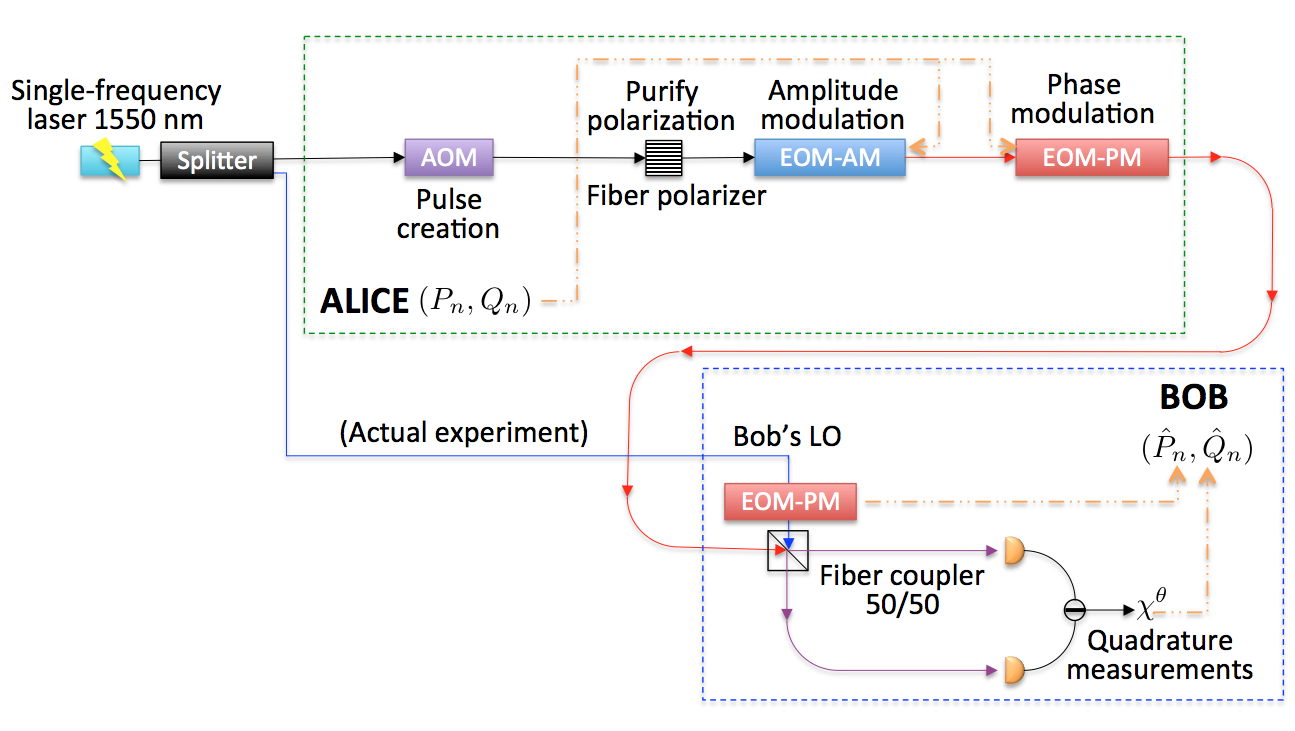}}
\caption{Schematic of experimental setup for proof-of-principle demonstration of SR-CV-QKD. The only difference between this schematic and Fig. \ref{fig:SR-CV-QKD} is the use of a shared laser source between Alice and Bob for experimental convenience.}
\label{fig:expt_schematic}
\end{figure}

The primary benefit of the SR-CV-QKD protocol is the reduction in hardware it enables at the transmitter and receiver. A schematic of our proof-of-principle experimental implementation of the SR-CV-QKD protocol is shown in Fig. \ref{fig:expt_schematic}. Since our purpose is to demonstrate the feasibility of the SR-CV-QKD protocol, the channel fiber length between Alice and Bob in all of the experiments reported on below was only 5~m.  Also, as shown in Fig. \ref{fig:expt_schematic}, we used a single laser for both Alice's and Bob's stations, which sat on the same optical table. We note that this was purely a matter of experimental convenience in a situation with limited resources, while any practical implementation of the protocol will definitely require separate laser sources for Alice and Bob. In addition, the use of a common laser source obviates the need to frequency lock Alice's and Bob's lasers, an additional experimental challenge that would have to be tackled in a practical setup as discussed in Sec. \ref{sec:SR-CV-QKD}. Nevertheless, crucial to this demonstration, due to the difference between the paths from the laser source to Alice's modulation component and from the laser source to Bob's detection apparatus, the phase difference between Alice's and Bob's frames was random and fluctuating (as shown below in Fig.~\ref{fig:data_session_pi}). 

Extending the SR-CV-QKD operation to practical distances and including the use of separate laser sources in Alice’s and Bob’s stations, are part of ongoing work in our laboratory. Further details of the current experimental setup are given in Appendix~\ref{app:expt_details}.

\begin{figure}[!tb]
\centering
\subfigure[~Reference pulses \label{fig:data_session_ref}]{\includegraphics[width=0.48\linewidth]{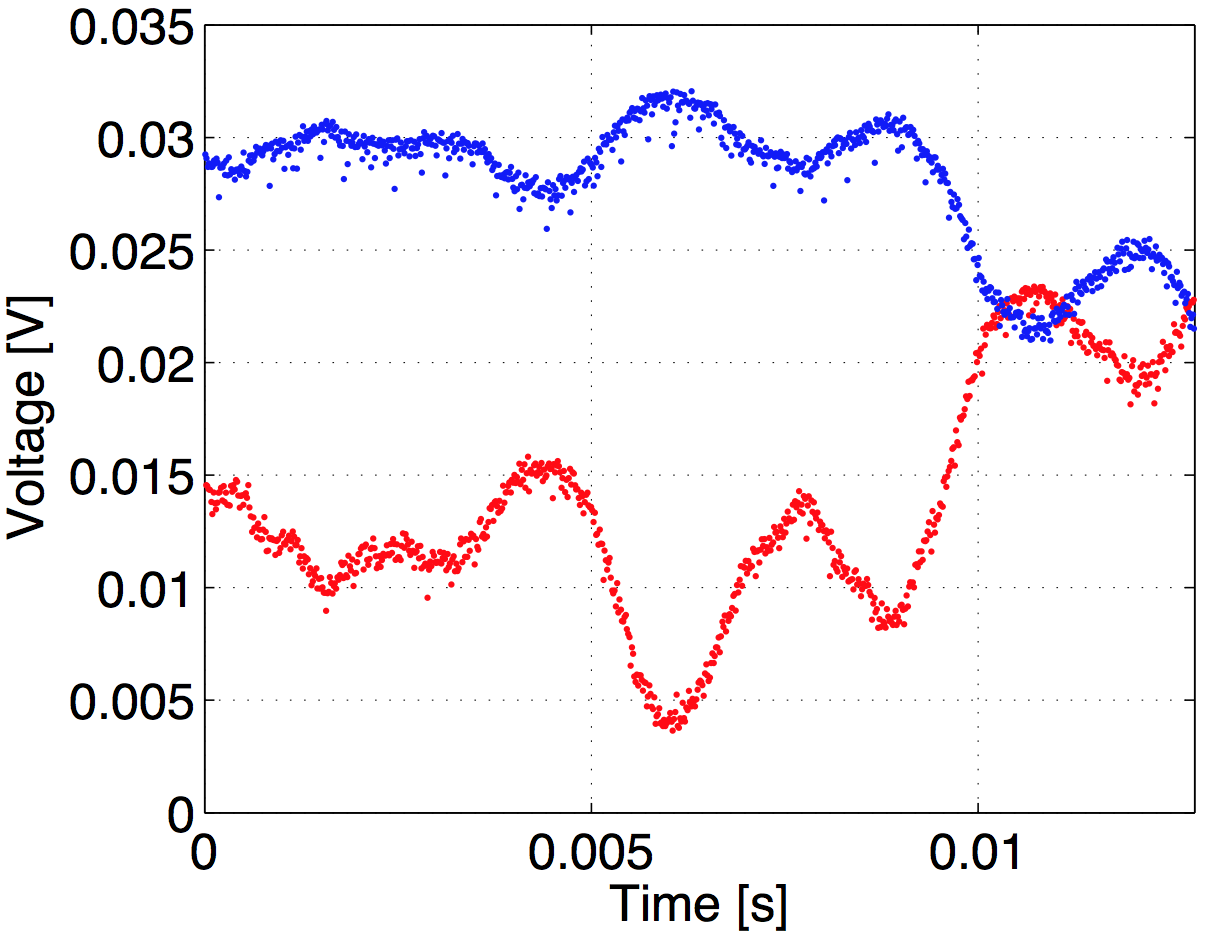}}
\subfigure[~Signal pulses \label{fig:data_session_sig}]{\includegraphics[width=0.48\linewidth]{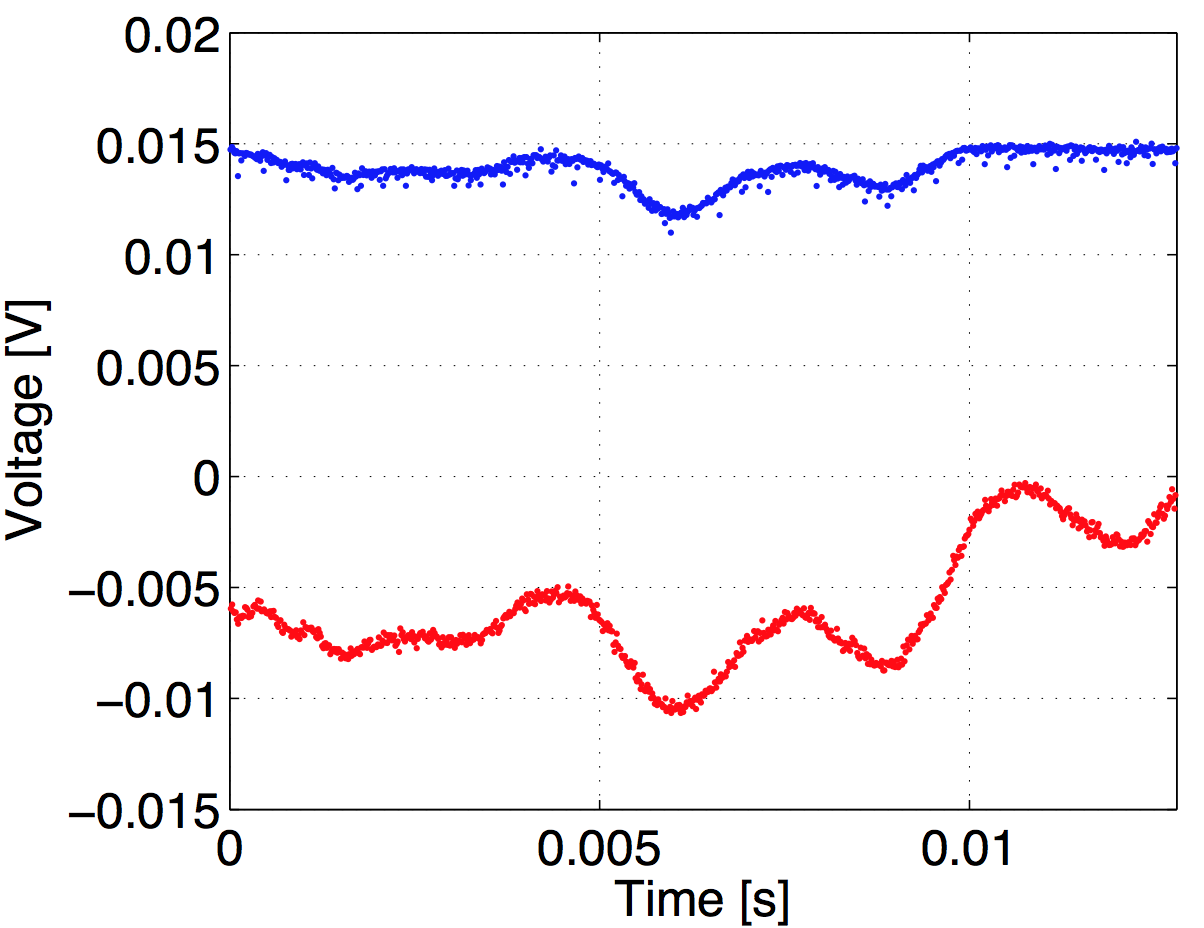}}
\subfigure[~Estimated phase angle \label{fig:data_session_est}]{\includegraphics[width=0.48\linewidth]{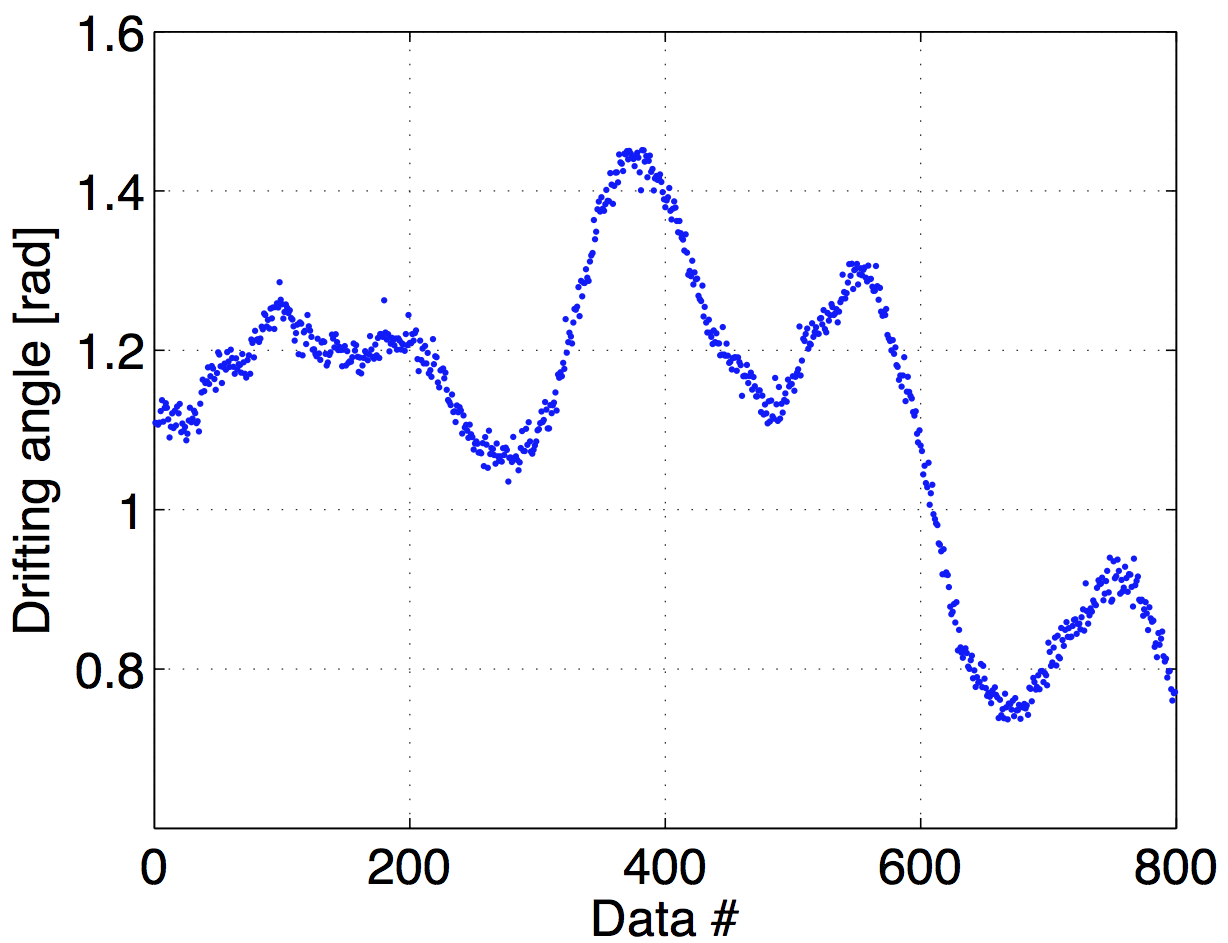}}
\subfigure[~Compensated signal \label{fig:data_session_comp}]{\includegraphics[width=0.48\linewidth]{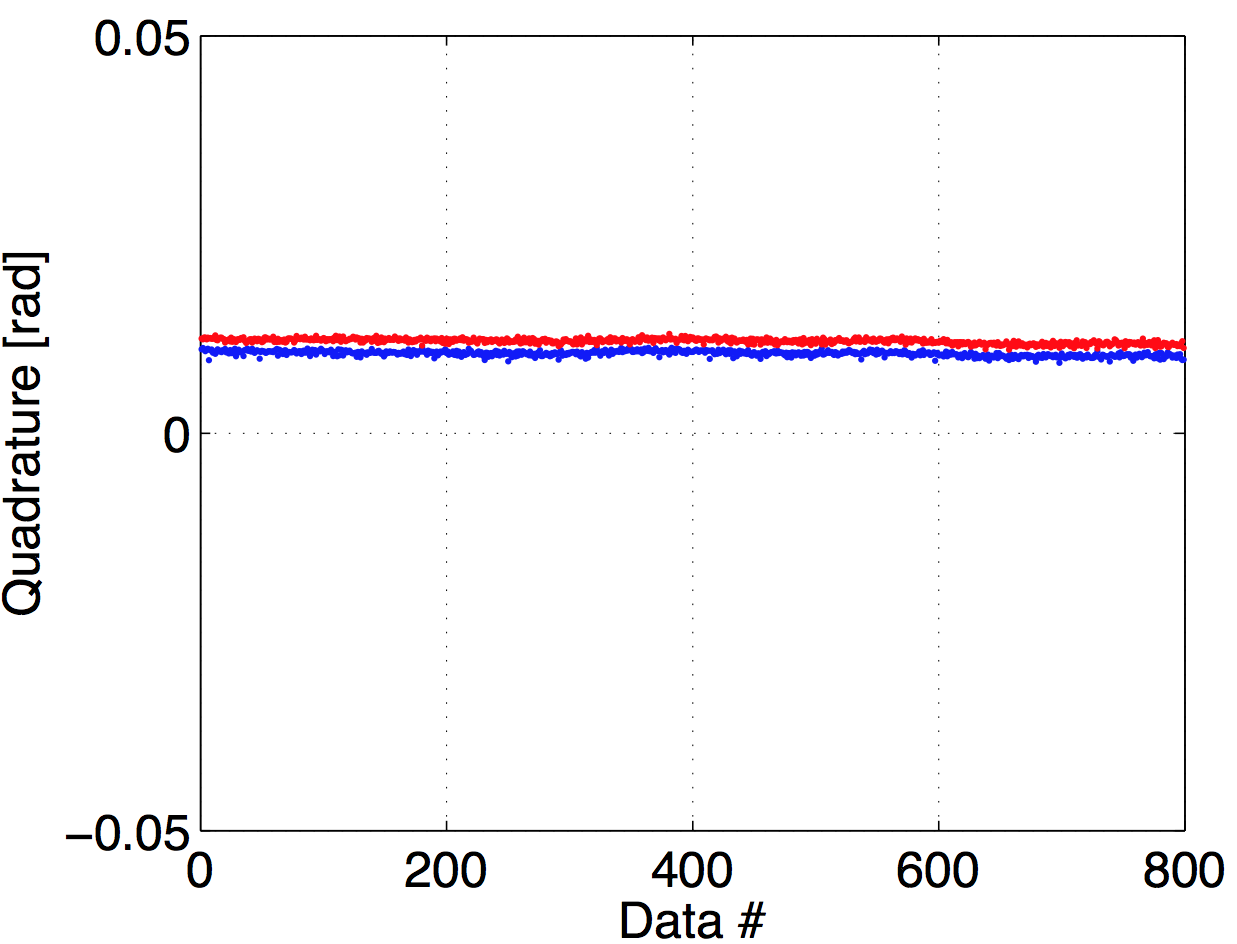}}
\caption{Phase drift estimation and compensation for constant signal pulses. Bob's measured voltages proportional to mean values of $Q$ quadrature (blue dots) and $P$ quadrature (red dots) for (a) reference pulses and (b) signal pulses. (c) Estimated values $\hat\theta$ of the phase difference between Alice's and Bob's frames. Each data point is calculated as in Eq.~\eqref{eq:thetahat_ratio_1} using mean quadrature values $q_{B_R}$ and $p_{B_R}$ obtained from sequential homodyne measurements on a pair of twin reference pulses. (d) Mean values of $Q$ quadrature (blue dots) and $P$ quadrature (red dots) for signal pulses, after a rotation by the angle $-\hat\theta$ to compensate for the phase drift.}
\label{fig:data_session_pi}
\end{figure}

\subsection{Signal retrieval under strong phase noise}
\label{sec:exp-1}

\begin{figure}[!tb]
\centering
\includegraphics[width=0.7\linewidth]{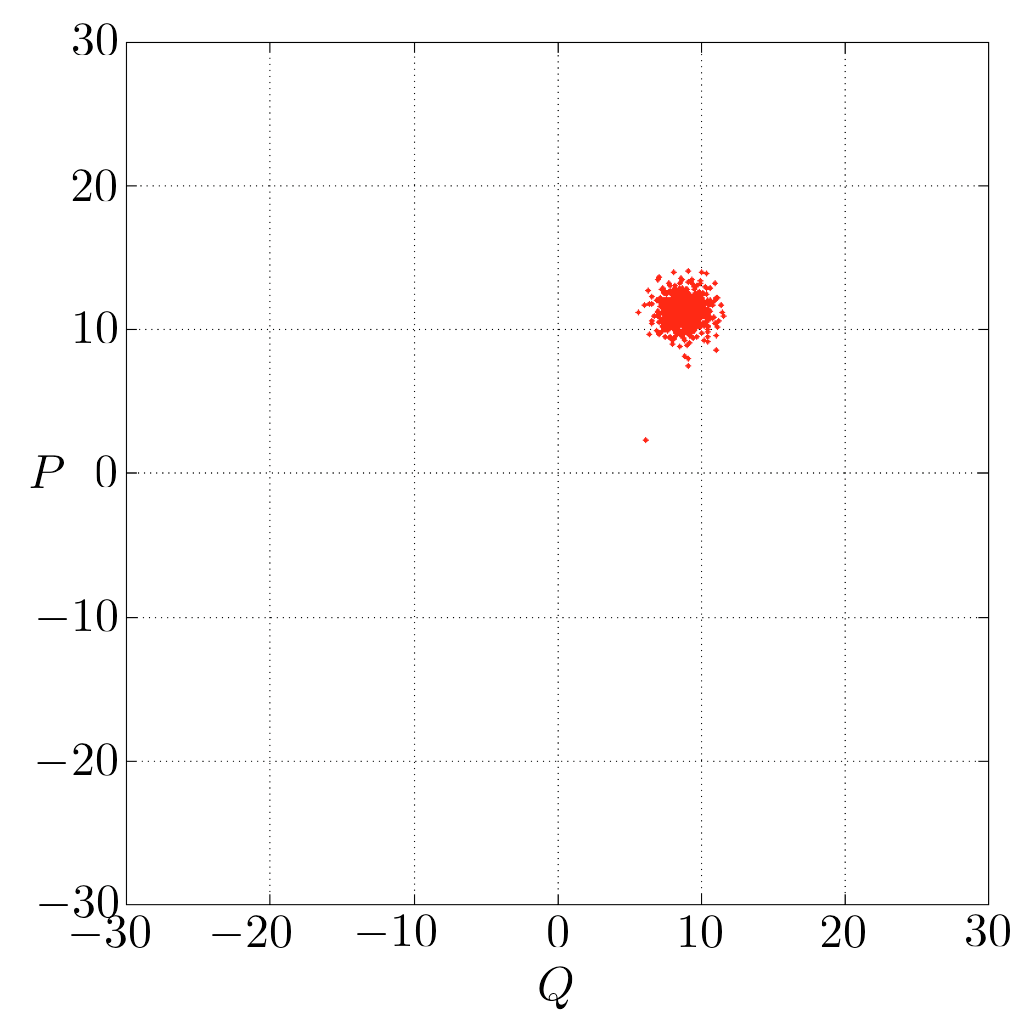}
\caption{The phase-space representation of reconstructed signal pulses after phase drift compensation.}
\label{fig:quantumstatereconstruction}
\end{figure}

We first demonstrate the ability of the protocol to detect the phase drift and compensate for it. For this purpose we let Bob compensate his measured quadrature values rather than send the phase estimate to Alice. Each signal pulse is prepared in a constant coherent state and is accompanied by a reference pulse that is also prepared in a constant coherent state with with mean quadrature values $(q_{A_R},p_{A_R}) = (30,0)$. Bob measures both $Q$ and $P$ quadratures of signal and reference pulses using his own LO, and the measured data are shown in Figs.~\ref{fig:data_session_ref} and \ref{fig:data_session_sig}. The fluctuations of measured mean quadrature values of both signal and reference pulses indicate the phase drift happening in the setup. Bob's estimate of the drifting phase, calculated using measurements of reference pulses' quadratures, is shown in Fig.~\ref{fig:data_session_est}, and mean quadrature values of signal pulses, compensated in accordance with the estimated phase, are shown in Fig.~\ref{fig:data_session_comp}. The compensation successfully recovers the constant signal that Alice sent despite the variation of the phase difference between Alice's and Bob's frames. The phase-space representation of Bob's compensated signal is shown in Fig.~\ref{fig:quantumstatereconstruction}. The variances of the reconstructed signal data is $1.16$.

We note that in this setup Bob was restricted to performing homodyne measurements and therefore could measure only one quadrature per pulse. Therefore Alice sent two identical signal pulses and two identical reference pulses in succession (the twin-pulse mode). The time taken for Bob to perform homodyne measurements on all four pulses (16~$\mu$s) is much shorter than the characteristic time of phase fluctuations ($f_{\theta}^{-1} \sim$ 200~$\mu$s).

\subsection{Signal constellation reconstruction and noise estimation}
 
\begin{figure}[!tb]
\centering
\includegraphics[width=0.8\linewidth]{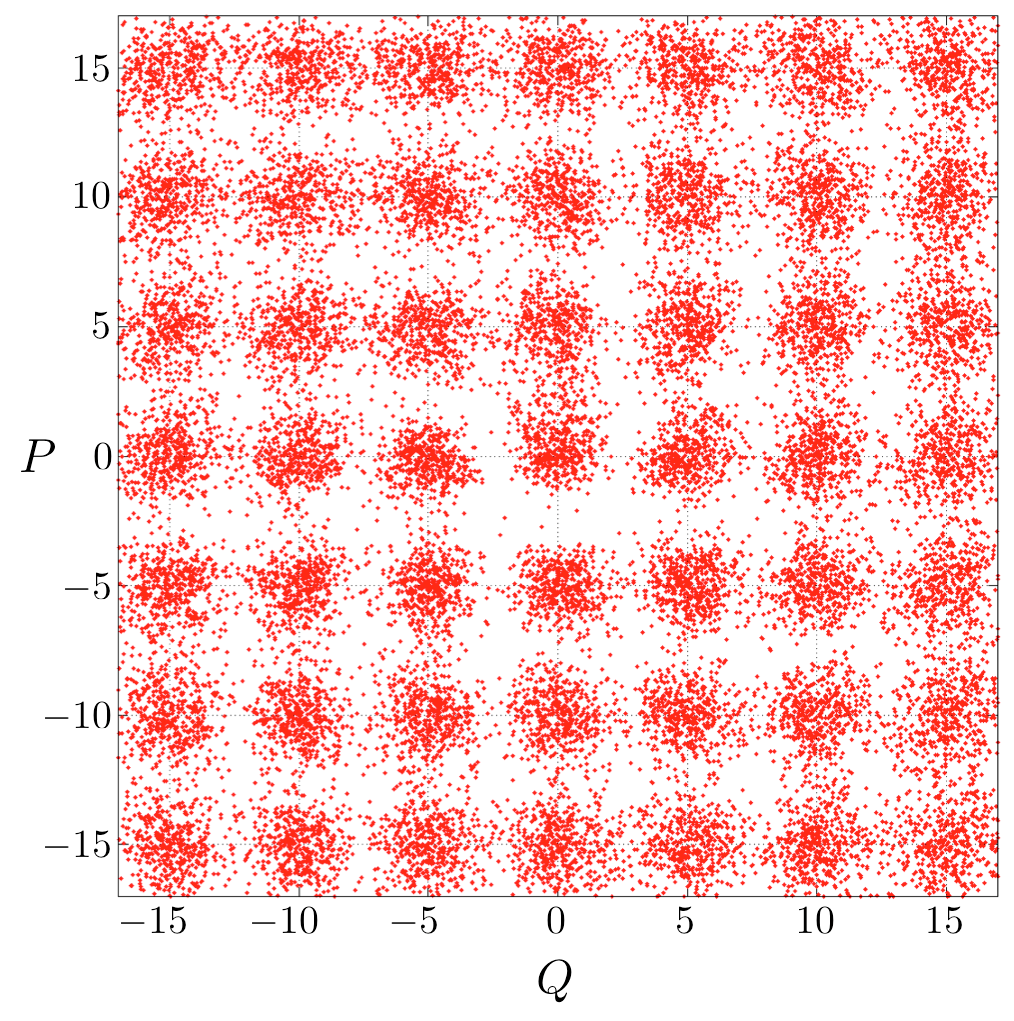}
\caption{The phase-space representation of signal pulses reconstructed using quadrature measurements on reference pulses and phase difference estimation.}
\label{fig:reconstruction} 
\end{figure}

Next, we report reconstruction of a constellation of signal pulses, which demonstrates the effectiveness of the self-referenced technique for signals over a large area in the phase space. In this experiment, we used a square grid in Alice's phase plane, with tiles of size $5 \times 5$ and centers of tiles spanning from -15 to 15 in both quadrature axes (for a total of 49 tiles). Alice generated 1000 identical signal pulses prepared in a coherent state centered on each tile, and sent them to Bob. Each pair of signal pulses was accompanied by a pair of reference pulses; all reference pulses were identical with mean quadrature values $(q_{A_R},p_{A_R}) = (30,0)$. Bob performed homodyne measurements of $Q$ and $P$ quadratures on each pair of twin signal pulses (resulting in 500 paired data points for each grid tile), as well as on each pair of twin reference pulses. In the same manner as described in Sec.~\ref{sec:exp-1} above, Bob used quadrature measurements on reference pulses to estimate the fluctuating phase difference between his and Alice's frames, and then compensated for estimated phase values to recover signal pulses sent by Alice. Figure~\ref{fig:reconstruction} shows the reconstructed constellation of signal pulses. The distribution of Bob's reconstructed signals is quite uniform in both quadratures for each of the grid tiles. However, one can notice the known ``zero anomaly'' \cite{usenko2010feasibility, shen2010experimental}, which is manifested as a skewed distribution of reconstructed values near the vacuum, arising due to a finite extinction ratio of Alice's electro-optic modulator (EOM) used for amplitude modulation. This undesirable effect can be mitigated by using an EOM with a higher extinction ratio or chaining multiple EOMs to achieve greater extinction. 

Typical Gaussian-modulated coherent-state pulses used as signals in CV-QKD have the modulation variance $V_A \sim 40$, and the purpose of the presented example is to demonstrate that such signals can be accurately reconstructed using reference pulses and phase difference estimation. We emphasize that during this entire experiment (which involved transmission of 49,000 signal pulses and reconstruction of 24,500 pairs of mean quadrature values) we did not have to concern ourselves with the stability of Bob's LO since the random phase drift was compensated at all times  using the self-referenced technique. 

\begin{figure}[!tb]
\centering
\includegraphics[width=0.8\linewidth]{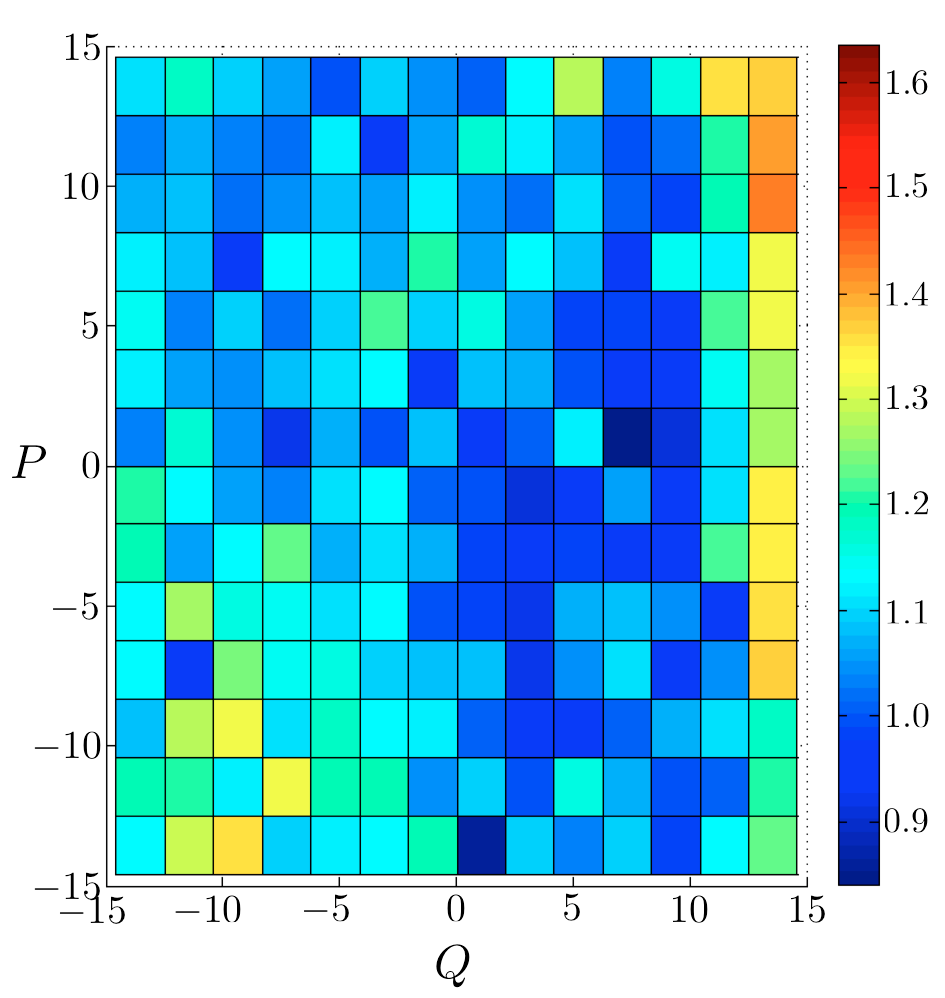}
\caption{The variance of the reconstructed signal data as a function of the phase-space location.}
\label{fig:variance_mapping_avg_1} 
\end{figure}

Next, we extended the characterization of signal pulses to the evaluation of the excess noise in the combined system, including Alice's encoding and Bob's decoding apparatus. In this experiment, we used the same reconstruction procedure for a constellation of coherent-state signal pulses as described above, but applied it to a different sample. With a finer grid tiling ($2.5 \times 2.5$), the number of tiles was four times greater (for a total of 196 tiles), but only 250 identical signal pulses were generated for each tile. After performing the reconstruction of the signal's mean quadrature values, we calculated the variance of the reconstructed data for each grid tile. Figure~\ref{fig:variance_mapping_avg_1} shows this variance as a function of the location on the phase-plane grid. Any variance in excess of $1$ is the excess noise introduced by the experimental apparatus, channel, or phase estimation procedure. We note that the variance distribution is quite uniform (0.95--1.2) over much of the considered phase-plane region. The average variance value over the entire constellation is 1.16. Thus the entire apparatus, including the phase compensation step, has a total excess noise of 0.16. This excess noise is due to several experimental imperfections, including non-uniform performance and calibration of EOMs across the phase plane and electronic noise in detectors.

\subsection{Demonstration of SR-CV-QKD with Gaussian modulation}

In addition to the above experiments that utilized the self-referenced technique for reconstruction of signal pulses in the presence of phase fluctuations and for characterization of the excess noise, we also performed the quantum components of an experimental secret key distribution using the SR-CV-QKD protocol. The QKD experiment was performed under a strong phase noise between Alice's signal-generating laser and Bob's LO.

We used a pseudo-random number generator based on the Mersenne twister for Alice's signal modulation and Bob's measurement axis selection. Each communication block consisted of 24,500 data points. Alice's pulse generation rate was 250 kHz; two thirds of these were identical reference pulses (used in the twin-reference-pulse mode), and one-third were signal pulses prepared in random Gaussian-modulated coherent states. Alice's Gaussian modulation variance was $V_A = 34$, and mean quadrature values of reference pulses were $(q_{A_R},p_{A_R}) = (30,0)$, corresponding to $V_R/V_A \approx 26.47$ \footnote{This (relatively low) value of the $V_R/V_A$ ratio is related to the fact that, in our proof-of-principle experiment, we placed more importance on the measurement precision rather than on the distance or rate of the key distribution. Specifically, due to the limited dynamic range of our analog-to-digital converter (ADC), the maximum value of $V_R$ had to be restricted in order to achieve a high precision of the signal homodyne measurement. This is not a fundamental restriction since ADCs with better resolution are available and moreover, in practical long-distance QKD, the required homodyne precision is less stringent (typically less than 5 binary digits), especially when the signal-to-noise ratio is less than 1.}. Since the transmission was only across an optical table, $T = 1$. Accounting for detector efficiency, homodyne visibility, and imperfections in homodyne arm balancing, Bob's homodyne efficiency was estimated to be $\eta = 0.719$ on average, however due to fluctuating mode-matching and homodyne arm balancing conditions this efficiency can fluctuate $\pm 0.1$.

At each round (consisting of two reference pulses and one signal pulse), Bob estimated the phase difference between his and Alise's frames and communicated this estimate to Alice, who performed the compensation rotation on the tabulated random values of the signal pulse quadratures.

In each session, among the 24,500 pulses exchanged, 2000 were randomly selected for estimation of the covariance matrix between Alice's and Bob's measurements during the parameter estimation stage. An example of such an estimated covariance matrix is:
\begin{equation}
	\overline{\gamma}_{AB}^{\mathrm{pm, exp}} = \left( \begin{array}{cccc} 
33.9637 & -0.9647 & 31.1408 & -0.0763 \\ 
-0.9647 & 34.3744 & -1.0830 & 30.7054 \\
31.1408 & -1.0830 & 29.4003 & -0.2307 \\ 
-0.0763 & 30.7054 & -0.2307 & 28.2540 \end{array}\right).
\end{equation}
Note that this covariance matrix corresponds to the prepare-and-measure version of the protocol and thus differs from $\overline\gamma_{AB}$ of Eq.~\eqref{eq:gamma_bar}. The theoretical form of $\overline{\gamma}_{AB}^{\mathrm{pm}}$ is
\begin{equation}
\overline{\gamma}_{AB}^{\mathrm{pm, the}} = \begin{pmatrix} 
V_A \openone & \sqrt{T \eta} V_A \overline{\cos \varphi} \openone \\ 
\sqrt{T \eta} V_A \overline{\cos \varphi} \openone & T \eta (V_A + 1 + \chi) \openone 
\end{pmatrix} ,
\end{equation}
where we used the assumption of symmetric $\mathcal{P}(\varphi)$ to set $\overline{\sin \varphi} = 0$. 
We can achieve reasonably good agreement between this theoretical form and the experimentally reconstructed covariance matrix using the above values for $V_A$, $V_R$, and $T$, along with the calibrated values $\epsilon = 0.01$ and $V_{\mathrm{el}} = 0.01$, and a value of homodyne efficiency at the upper limits of our calibrated range, $\eta=0.8$, which yields the theoretical covariance matrix
\begin{equation}
	\overline{\gamma}_{AB}^{\mathrm{pm, the}} = \left( \begin{array}{cccc} 
34.000 &  0.000 & 30.389 &  0.000 \\ 
 0.000 & 34.000 &  0.000 & 30.389 \\
30.389 &  0.000 & 28.218 &  0.000 \\ 
 0.000 & 30.389 &  0.000 & 28.218 \end{array}\right).
\end{equation}

Using the above parameter values, we also calculate mutual information bounds: $I_{A' B} \gtrapprox 2.37$~bit/round, $I_{E B} \leq 1.492$~bit/round, and $\chi_{BE} \lessapprox 1.742$~bit/round (a round consists of three pulses, two reference and one signal). Assuming the reconciliation efficiency value of $\beta = 0.95$, the expected minimum key rates secure against individual and collective attacks are $K_{\mathrm{ind}}^{\mathrm{min}} \approx 0.759$~bit/round and $K_{\mathrm{col}}^{\mathrm{min}} \approx 0.509$~bit/round, respectively. Taking into account that SR-CV-QKD in the twin-reference-pulse mode utilizes three pulses per round, the key rates per physical pulse would be three times lower. Finally, with the pulse generation rate of 250 kHz, the expected minimum key rates are $K_{\mathrm{ind}}^{\mathrm{min}} \approx 63.26$~kbit/s and $K_{\mathrm{col}}^{\mathrm{min}} \approx 42.45$~kbit/s. 

It should be noted that the primary factor that dictates the key rate in our setup is the pulse generation rate. We had to maintain this at the low rate of 250 kHz because the data acquisition hardware we used (National Instruments, PCIE-6363) supports only 250 kHz signal generation and measurement rates (taking into account rise and fall times of pulses). 

\begin{figure}[!tb]
\centering
\includegraphics[width=1.05\linewidth]{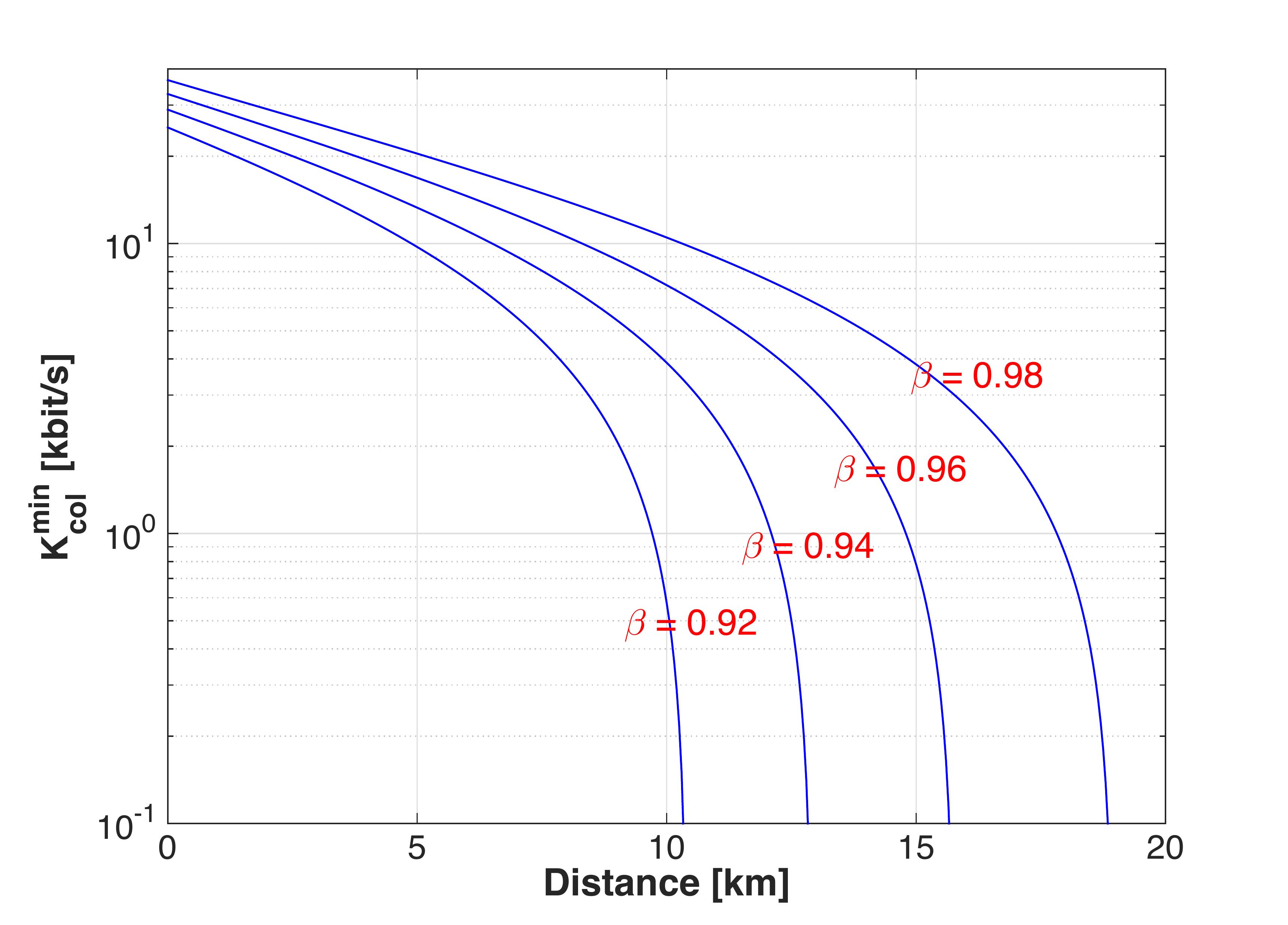}
\caption{Expected key rates as a function of transmission distance for current experimental setup and a range of post-processing efficiency values ($\beta$). The parameters used are: $V_A = 34, V_R = 900, \delta_R=0, \epsilon=0.01, V_{\rm el}=0.01$ and pulse rate of 250 kHz. The homodyne efficiency was taken to be the average value (this quantity fluctuates in our setup --- see main text) of $\eta=0.719$. The transmission loss was taken to be $0.2$ dB/km. }
\label{fig:expt_key_rates} 
\end{figure}

This demonstration is a proof-of-principle of the SR-CV-QKD protocol and the feasibility of phase difference estimation and compensation using reference pulses. Therefore we choose a negligible transmission distance. In Figure \ref{fig:expt_key_rates} we plot the achievable key rates as a function of transmission distance using this experimental setup for a range of post-processing efficiencies. The next generation of this experiment will include an upgrade of this hardware to increase pulse generation rates and focus on increasing key rate and extending key distribution distance using SR-CV-QKD.

\section{Conclusions and Discussion}
\label{sec:conc}

We have developed a new protocol, SR-CV-QKD, that eliminates the need to transmit an LO. The removal of this demand dramatically simplifies the hardware required to perform CV-QKD and removes the most significant obstacles to developing integrated photonics implementations of CV-QKD transceivers. We thus believe that this new protocol will play a key role in enabling the miniaturization of CV-QKD hardware, which has the potential to significantly enhance the applicability of quantum communications.

In the reported experiments, we characterized the core new element of SR-CV-QKD, signal reconstruction through compensation of the drifting phase, and performed a proof-of-principle demonstration of key distribution using the new protocol. On the theory side, we computed expected key rates, secure under a passive Gaussian channel assumption. A principal feature of our security analysis is the incorporation of the inherent quantum uncertainty of reference pulses. We showed that as the reference pulse amplitude increases, the key rate of SR-CV-QKD approaches that of conventional CV-QKD with LO transmission, but that this rate can be achieved with much simpler hardware. Our analysis has focused on asymptotic key rates as a first step in understanding the new prototol, but we expect that recent results in that calculate secure key rates with finite-size effects included \cite{Furrer:2012dz, Furrer:2014dz, Leverrier:2015tr} can be adapted to SR-CV-QKD.

A way to view the difference between SR-CV-QKD and conventional CV-QKD is that while the latter physically transmits a reference frame (in the form of the LO), the former only transmits \emph{information} about the reference frame. As a result, SR-CV-QKD is immune against many of the recently identified side-channel attacks that exploit detection using a publicly shared high-power LO \cite{Jouguet.PRA.87.062313.2013, Huang.PRA.87.062329.2013}. Of course, it remains to be seen whether new side channel attacks that target SR-CV-QKD are possible.

In this work, we focused on a version of SR-CV-QKD, in which Alice prepares signal and reference pulses in coherent states, and Bob performs a homodyne measurement on the signal pulse. More generally, it is possible (in analogy with a variety of conventional CV-QKD protocols~\cite{GarciaPatron:2007ws}) to consider alternative versions of SR-CV-QKD. For example, Alice can prepare the signal pulse in a squeezed state, Alice can prepare a pair of reference pulses in orthogonal squeezed states, Bob can perform a heterodyne measurement on the signal pulse, Bob can perform a homodyne measurement of a random quadrature, and so on.

It should be noted that in classical optical communication based on coherent detection, reference laser pulses have been used as a direct phase reference for signal generation and phase noise cancellation and stabilization.  In particular, in formats involving quadrature phase-shift keying and quadrature amplitude modulation, the self-homodyne detection (SHD) approach~\cite{Miyazaki:2005jq, Sjodin:2011hg, Shinada:WHGPx0LX} makes use of dedicated pilot carrier pulses that are multiplexed with the signal pulses from the same laser to enable homodyne detection at the receiver. SR-CV-QKD is in the same spirit as these techniques, except that, unlike SHD, our protocol enables absolute amplitude and phase measurement in the low-photon-number regime with encoding and decoding capabilities across a continuous amplitude and phase variation in real-time. The method of phase estimation and compensation underlying the SR-CV-QKD protocol can be considered a close quantum analog of intradyne detection \cite{Nakazawa:2010ek}, another approach used in classical coherent communication, which digitally estimates phase (and frequency) drifts using measurements on part of the modulated signal pulse. An important aspect of adapting this technique to CV-QKD is that the measurement for phase estimation should be performed on a dedicated reference pulse rather than directly on the signal pulse, because at the intensity levels used for CV-QKD, detection of any portion of the signal pulse would severely decrease achievable key rates. As a matter of fact, the precision of phase estimation achieved via the direct measurement on the signal pulse would correspond to SR-CV-QKD with $V_R/V_A < 1$. According to Fig. \ref{fig:keyrates}, such a low $V_R/V_A$ value would not be practical (this is because QKD operates near shot-noise levels). Therefore, dedicated reference pulses with sufficiently large amplitude are required as prescribed by SR-CV-QKD.

Previous CV-QKD experiments have made use of strong calibration pulses to compensate for phase drifts created by the signal and LO having different propagation paths at the receiver \cite{Qi:2007ca,Wittmann:2010hh,Khan:2013jw}. However, it should be noted that these experiments co-transmitted the signal and LO and hence do not use the phase compensation to its full extent. 
Also, in the context of the B92 DV-QKD protocol \cite{Bennett:1992zz}, Koashi has constructed a modified scheme whereby Bob estimates the phase difference between his and Alice's lasers \cite{Koashi:2004hf}. However, unlike in SR-CV-QKD where this estimate is used to modify Alice's classical data, Koashi's scheme uses the estimate to phase shift a weak field that interferes with Alice's signal. The SR-CV-QKD protocol is practically much simpler since no dynamic tuning of optical components conditioned on the phase estimate (optical feed forward control) has to be performed.

We note that this new CV-QKD protocol was independently discovered by Qi~\etal, as recently reported in Ref.~\cite{Qi:2015uf}. Qi~\etal~present a complementary study of the protocol, including its implementation to perform key distribution over a 25~km link, which goes beyond our proof-of-principle demonstration. In contrast, in this work, we focused on a comprehensive analysis of the fundamental limits of the protocol (expected secret key rate calculations taking into account the quantum uncertainty of reference pulses and accuracy of the phase estimation, in Sec.~\ref{sec:theory}), and characterization of the performance of the central new element in the protocol: phase drift compensation using reference pulses (in Sec.~\ref{sec:experiment}). Note that the expected key rates calculated by Qi~\etal are larger than those calculated in this work because while we use the strictest security criterion \cite{Scarani:2009wq}, Qi~\etal use the relaxed criterion where calibrated noise and loss at Bob's receiver are assumed to be out of Eve's control. Our results, along with the demonstration in Qi~\etal~\cite{Qi:2015uf}, establish SR-CV-QKD as a practical protocol with significant benefits in terms of hardware simplification and potential compatibility with integrated photonics.

Finally, while this manuscript was under review Huang et al. reported on an implementation of CV-QKD over a 25-km link without transmission of a local oscillator, which utilizes a protocol that is essentially the same as SR-CV-QKD \cite{Huang:2015dy}.

Ongoing work in our laboratory is focused on increasing the key rate and transmission distance of the SR-CV-QKD link and increasing the stability and robustness of the transmitter and receiver components.
 
\acknowledgments
We are grateful to Chris DeRose (SNL), Paul Davids (SNL), Tony Lentine (SNL) and Christian Weedbrook for informative discussions about integrated photonics and CV-QKD.
This work was supported by the Laboratory Directed Research and Development program at Sandia National Laboratories. Sandia is a multi-program laboratory managed and operated  by Sandia Corporation, a wholly owned subsidiary of Lockheed Martin Corporation, for the United States Department of Energy's National Nuclear Security Administration under contract DE-AC04-94AL85000.

\appendix
\section{Experimental details}
\label{app:expt_details}

In this appendix, we provide the details of our experimental setup. 

The laser source was a fiber-pigtailed New Focus single-frequency laser at 1550~nm, with an optical bandwidth of $\sim 100$~kHz, and a maximum output power of 7~mW. After an in-line polarizer, the polarization extinction ratio was 35~dB (all fiber used in the experiment was polarization maintaining Panda PM1300 fiber). An acousto-optic modulator (Brimrose) modulated the output of the laser at 250~kHz with individual pulse duration of 200~ns. The amplitude and phase of the light were modulated through 10~GHz fiber-pigtailed amplitude and phase EOMs (Thorlabs).

At Bob's station, received light pulses were detected using a homodyne setup, as shown in Fig.~\ref{fig:SR-CV-QKD}. Although in practice Bob will use an independent laser as an LO for homodyne measurements, in our experiment, for simplicity, Alice and Bob shared the same single-frequency laser source. However, it is critical to note that, due to the difference between the paths from the laser source to Alice's modulation component and from the laser source to Bob's detection apparatus, the phase difference between Alice's and Bob's frames was random and fluctuating (as shown in Fig.~\ref{fig:data_session_pi}). Bob's LO power was 0.1~mW. For homodyne detection, we used a polarization-maintaining fiber beam splitter with an approximately 51:49 splitting ratio (Thorlabs). We attached a mechanical variable optical attenuator to each leg after the beam splitter to balance the power between the two legs. For the balanced detector, we used a commercially available switchable-gain fast InGaAs detector (Thorlabs PDB450C). The detection bandwidth was set at 45~MHz where we observed that the dark current noise was approximately 25~dB lower than the shot noise from Bob's LO.  Accounting for detector efficiency, homodyne visibility, and imperfections in homodyne arm balancing, Bob's homodyne efficiency was estimated to be $\eta = 0.719$ on average, however due to fluctuating mode-matching and homodyne arm balancing conditions this efficiency can fluctuate $\pm 0.1$. We are currently making progress in understanding the root causes of this uncertainty and how to stabilize it. 

For data collection and analog voltage generation for the EOMs, we used a commercially available data acquisition card (NI PCIE-6363), capable of reliably collecting multi-channel data at 250~kHz. Since the analog output of this card has 1~M$\Omega$ impedance while the EOM's RF modulation input has 50~$\Omega$ impedance, we built in a fast unity-gain voltage follower to match the impedances. All data generation and collection were performed through Matlab's data acquisition toolbox based codes.

\section{Device calibration with reference pulses}
\label{sec:calibration}

The technique of using reference pulses and phase estimation to compensate for phase drifts is valuable not only for running the CV-QKD protocol \emph{per se}, but also for the calibration of Alice's and Bob's apparatus. Since CV-QKD operates at the limits of detection, it is vital to calibrate, and maintain calibration of, the modulators and homodyne detectors in the setup. Typically, this calibration is done locally by Alice and Bob to minimize security loopholes, and for this reason Alice's station should have homodyne or heterodyne detection capabilities and Bob's station should have pulse generation capabilities. In this appendix, we present a calibration task that Alice and Bob need to perform and show that it benefits greatly from the use of reference pulses and phase estimation.

\begin{figure}[t]
\centering
\subfigure[~Raw data \label{fig:calib_raw}]{\includegraphics[width=0.481\linewidth]{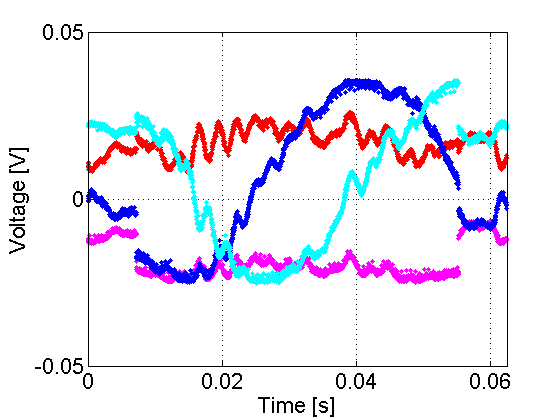}}
\subfigure[~Calibration curve after phase drift compensation \label{fig:calib_comp}]{\includegraphics[width=0.48\linewidth]{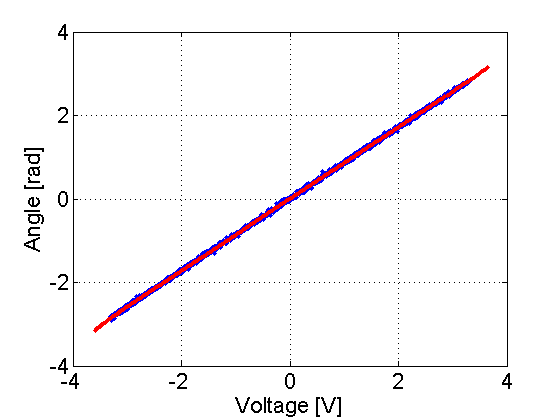}}
\caption{In-situ calibration of a phase EOM using reference pulses and phase drift compensation. (a) Raw data collected while the modulation voltage is swept from $-3.5$~V to $3.5$~V within the time window (5--58~ms); red:  in-phase value of the reference pulse, magenta: quadrature value of the reference pulse, blue: in-phase value of the phase-modulated pulse, cyan: quadrature value of the phase-modulated pulse. (b) The calibration curve between the applied voltage and the induced phase after phase drift compensation; blue dots: calibration data, red line: polynomial fit.}
\label{fig:eom-calibration}
\end{figure}

It is critical that the phase induced by Alice's and Bob's phase EOM is well calibrated against the applied voltage; i.e., the actual phase modulation should correspond accurately to the random numbers generated by Alice, or the measurement axis chosen by Bob. In a long-running CV-QKD implementation, this calibration may have to be performed repeatedly since EOM characteristics can drift over time. The calibration requires performing a test phase modulation and measuring its value. The measurement is performed using a homodyne setup that utilizes an LO generated from the same master laser as the modulated pulse. As a result, this calibration requires precise knowledge of the path difference (which results in a relative phase shift) between the modulation path and the LO path. This path difference can fluctuate due to thermal effects and tracking it requires a considerable effort.

Noting that the calibration problem in the presence of a phase drift is very similar to the problem of establishing a common phase reference in CV-QKD, it is clear that we can alternatively use reference pulses to perform the EOM phase calibration. That is, during the calibration stage, each phase-modulated pulse is accompanied by a reference pulse (with no phase modulation). The procedure of heterodyne measurement and phase estimation on this reference pulse tells us what the reference zero phase modulation value is, and this can be used to recover the actual phase of the modulated pulse.

Figure~\ref{fig:calib_raw} shows raw measured values for the phase-modulated pulse when the EOM voltage is swept from $-3.5$~V to $3.5$~V, along with those for the accompanying reference pulse. Within this voltage range the EOM response is reasonably linear. The raw voltages show that the phase does drift over the timescale of the sweep. However, Fig.~\ref{fig:calib_comp} shows that by estimating the drifting phase and compensating for it, one can obtain a clean calibration map between the applied voltage and the induced phase.

\bibliography{nlo_cv_qkd}

\end{document}